\documentclass[aps,prl,reprint,superscriptaddress,amsmath,amssymb,longbibliography]{revtex4-2}

\usepackage{graphicx}
\usepackage{bm}
\usepackage{booktabs}
\usepackage{xcolor}
\usepackage[colorlinks=true,linkcolor=blue!60!black,citecolor=blue!60!black,urlcolor=blue!60!black]{hyperref}

\newcommand{\dn}{\mathrm{d}}

\makeatletter
\providecommand{\startappendixtoc}{%
  \let\l@section\appendixtocsection}
\providecommand{\stopappendixtoc}{%
  \let\l@section\@gobbletwo}
\makeatother

\begin{document}

\title{Chaotic singularities inside vacuum black holes} 

\author{Robie A.~Hennigar}
\email{robie.a.hennigar@durham.ac.uk}
\affiliation{Centre for Particle Theory, Department of Mathematical Sciences, Durham University, Durham DH1 3LE, UK}
\author{Ayan K.~Patra}
\email{ayan-kumar.patra@durham.ac.uk}
\affiliation{Centre for Particle Theory, Department of Mathematical Sciences, Durham University, Durham DH1 3LE, UK}

\begin{abstract}
According to the conjecture of Belinski, Khalatnikov and Lifshitz, general relativity gives rise to chaotic dynamics in the vicinity of generic spacelike singularities. However, no vacuum black hole has been shown to exhibit this behaviour. Here we construct stationary, asymptotically locally AdS$_5$ black holes with regular, triaxially squashed horizons and causal exteriors, whose homogeneous interiors realize the cosmological billiard of generic five-dimensional vacuum gravity. We identify a simple algebraic condition on the horizon data that is necessary for the interior billiard to be chaotic, and show that solutions with regular horizons locally access an open set of billiard trajectories. 
\end{abstract}

\maketitle

\emph{Introduction.}---The theorems of Penrose and Hawking establish that singularities are generic in general relativity, but say little about the behavior of spacetime in their vicinity. For spacelike singularities, a compelling dynamical picture is provided by the conjecture of Belinski, Khalatnikov and Lifshitz (BKL)~\cite{Belinsky:1970ew}. In four-dimensional vacuum gravity, BKL argued that as a generic spacelike singularity is approached, neighbouring points decouple, and each point evolves through an infinite, chaotic sequence of Kasner epochs. Subsequent work extended the BKL conjecture to higher dimensions and mapped the near-singularity dynamics to particle motion in a hyperbolic billiard~\cite{Chitre:1972,Misner:1994ge, Damour:2001sa, Damour:2002et}. The billiard walls are determined by the field content of the theory, and the dynamics is chaotic precisely when the resulting chamber has finite volume. For pure gravity, this occurs for $D \le 10$~\cite{Demaret:1985jnc,Demaret:1988sg,Damour:2001sa}.

The BKL analysis is local, so it should apply equally well for the initial singularities of cosmology and the spacelike portions of final singularities inside black holes. In cosmology, the conjecture is supported by numerical simulations of inhomogeneous models~\cite{Berger:1998vxa,Garfinkle:2003bb} and by rigorous results for homogeneous ones~\cite{Ringstrom:2000mk,Heinzle:2007kv,Heinzle:2009eh}. Much less is known about black holes, where the cosmological results do not immediately carry over. 
A black hole interior can be viewed as a collapsing cosmology, but it must continue through a regular horizon into a physically acceptable exterior. These requirements impose additional constraints that may be incompatible with oscillatory singularities.  

A useful starting point is to consider black holes with spatially homogeneous interiors. Homogeneity reduces the Einstein equations to ordinary differential equations, which makes the problem more transparent and tractable. In this context, explicit examples of black holes realizing BKL chaos have been obtained in asymptotically AdS spacetimes by coupling Einstein gravity to a collection of massive vector fields~\cite{DeClerck:2023fax,Caceres:2026mug}.  Near the singularity, the vector fields supply electric walls that enclose a finite-volume billiard and drive transitions between Kasner epochs.  To the best of our knowledge, these constructions provide the only examples of BKL dynamics inside black holes, although the choice of matter sector introduces subtleties in their physical interpretation.\footnote{These examples use negative squared mass for the vector to ensure absence of inner horizons. The models satisfy the AdS BF bound but violate the CFT unitarity bound~\cite{DeClerck:2023fax,Caceres:2026mug}. It is expected that extensions exist that avoid this problem~\cite{DeClerck:2023fax}, but these have not yet been constructed. A further qualification concerns longitudinal modes, which the ansatz suppresses. For positive mass squared, exciting these modes can open the billiard and terminate the oscillations~\cite{Henneaux:2022ijt}.} 

The situation in vacuum gravity is more restrictive. In four dimensions, the natural candidate in vacuum is the Bianchi~IX (Mixmaster) universe, whose chaotic evolution is driven by the curvature of its spatial slices~\cite{Misner:1969hg}. However, it admits a regular horizon only if two of its scale factors coincide, which eliminates the chaos~\cite{Siklos:1978,Chrusciel:1994be,Ringstrom:1999hy, ForthcomingBKL}. This strongly suggests that homogeneous vacuum black holes in four dimensions cannot have chaotic BKL interiors,  and whether vacuum gravity can support such a black hole at all has remained an open question.

Here we answer this question in five dimensions. We construct vacuum, asymptotically locally AdS$_5$ black holes whose interiors realize the full billiard of generic five-dimensional gravity. The solutions are stationary and realize an $\mathbb{R}\times SU(2)$ isometry group. Their conformal boundary is a timelike fibration over a squashed three-sphere. The key difference relative to four dimensions is that here the horizon generator commutes with $SU(2)$, allowing the horizon to be fully triaxial. We identify how the horizon data determine the surviving billiard walls and verify that the limiting cosmological billiard accurately describes the dynamics. We further find that black holes with causal exteriors access an open set of billiard positions and trajectories.  

\emph{Setup.}---We consider Einstein gravity with $R_{\mu\nu}=-4L^{-2}g_{\mu\nu}$ and metrics of the form
\begin{equation}
\dn s^2=\dn\rho^2-F^2\bigl(\dn T+A_a\sigma^a\bigr)^2+h_{ab}\,\sigma^a\sigma^b ,
\label{eq:ansatz}
\end{equation}
where $\sigma^a$ are the left-invariant one-forms on $SU(2)\simeq S^3$, with $\dn\sigma^a=\tfrac12\epsilon_{abc}\sigma^b\wedge\sigma^c$. The ansatz is characterized by a function $F$, Kaluza-Klein connection $A_a$, and the symmetric matrix $h_{ab}$, all of which depend only on $\rho$. The isometry group is generically $\mathbb R\times SU(2)$, and the Einstein equations reduce to ordinary differential equations for ten functions. The ansatz contains Schwarzschild--AdS$_5$ ($h_{ab}\propto\delta_{ab}$, $A_a=0$), static squashed black holes~\cite{Murata:2009jt, Brihaye:2009dm}, and the equal-spinning Myers--Perry--AdS$_5$ black hole~\cite{Hawking:1998kw}, each of which has an additional Killing field. Black resonators~\cite{Ishii:2018oms} also share the same $\mathbb{R}\times SU(2)$ symmetry, but with the horizon generator $k = \partial_T$ becoming spacelike near the boundary. In all of these cases, the connection either vanishes or lies along a principal axis of $h_{ab}$. Here we study the general case, with $A_a\neq0$ and nondiagonal $h_{ab}$, but we shall impose that $k$ is timelike throughout the exterior. 

We seek solutions with regular horizons, which must lie at zeroes of $F$. Smooth extension across the horizon requires $F=\kappa\rho+O(\rho^3)$, $h_{ab}=H_{ab}+O(\rho^2)$ and $A_a=A^{\rm H}_a+O(\rho^2)$, with all metric components smooth in~$\rho^2$. The near-horizon expansion is then fully determined by $\kappa$, the horizon metric $H_{ab}$, and the horizon connection $A^{\rm H}_a$. By rescaling time and applying constant $SO(3)$ rotations to $\sigma^a$, we can set $\kappa=1$ and diagonalize $H_{ab}$~\footnote{Note that diagonalizing $H_{ab}$ at the horizon does not mean $h_{ab}$ will be diagonal off the horizon.}. The solutions are therefore specified by six physical parameters: the eigenvalues $H_1,H_2,H_3$ of $H_{ab}$, and the components of $A^{\rm H}_a$ in that eigenbasis. Each horizon cross section is a homogeneous, (generically) triaxially squashed three-sphere. 

Near the conformal boundary, $z=e^{-\rho/L}$ puts Eq.~\eqref{eq:ansatz} in Fefferman--Graham form. Rescaling time, rotating the $\sigma^a$, and performing a constant Weyl transformation brings the boundary metric to the canonical form
\begin{align}
\dn s^2_{(0)}={}&-\bigl(\dn\hat T+b_1\sigma^1+b_2\sqrt{r_2}\,\sigma^2+b_3\sqrt{r_3}\,\sigma^3\bigr)^2\nonumber\\
&+(\sigma^1)^2+r_2(\sigma^2)^2+r_3(\sigma^3)^2 ,
\label{eq:boundary}
\end{align}
with $1\le r_2\le r_3$, and $b_i$ the components of the boundary connection in an orthonormal frame. Both bulk and boundary are free of closed causal curves if $v^2\equiv F^2h^{ab}A_aA_b<1$. The boundary data consist of the five constant parameters of the boundary metric, along with six independent components of the holographic stress tensor. The six horizon parameters therefore suggest one-parameter families of black holes at fixed boundary geometry, labelled by temperature.

\begin{figure}[t]
\centering
\includegraphics[width=\columnwidth]{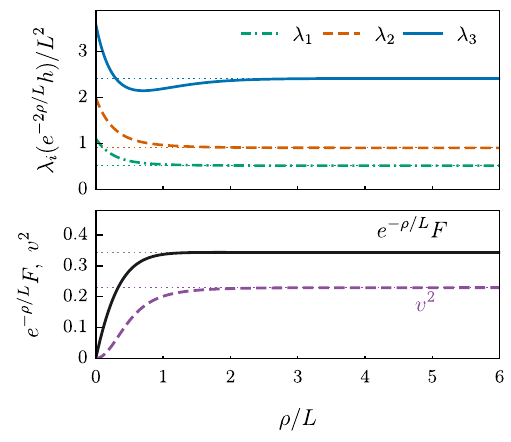}
\caption{Black hole exterior with horizon data in
Eq.~\eqref{eq:example-horizon}, in units $L=\kappa=1$. Upper panel: ordered eigenvalues of the full matrix $e^{-2\rho/L}h_{ab}$, in units of $L^2$. Lower panel: $e^{-\rho/L}F$ (black) and $v^2=F^2h^{ab}A_aA_b$ (purple). Dotted lines indicate the limiting boundary values. }
\label{fig:exterior-solution}
\end{figure}

We construct black holes numerically by integrating outward from the near-horizon solution. We show an explicit example in Fig.~\ref{fig:exterior-solution}. Here we have set $L = \kappa = 1$ and chosen 
\begin{align}
H_{ab}&=2\left(\frac{399}{400}\right)^{1/3}
\operatorname{diag}\!\left(\frac95,1,\frac59\right)\,\nonumber\\
A^{\rm H}_a&=\left(\frac{24}{25},\frac{12}{25},\frac{18}{25}\right)\,.
\label{eq:example-horizon}
\end{align}
The rescaled metric functions approach finite values at the boundary, with $F > 0$ and $h_{ab}$ positive definite throughout the exterior. The solution is causal, with $v^2$ increasing from zero at the horizon to approximately
$0.2294$ at the boundary. Since $g^{TT}=-(1-v^2)/F^2<0$, the coordinate $T$ provides a time function outside the horizon. After transforming this solution into the canonical form of Eq.~\eqref{eq:boundary}, the sources are
$(r_2,r_3)\simeq(1.744,4.642)$ and
$(b_1,b_2,b_3)\simeq(-0.3505,0.1836,0.2698)$.

\emph{The interior as a billiard.}---The black hole interior is obtained by continuing the solution through the horizon $\rho^2 \to - t^2$, giving an interior metric $\dn s^2=-\dn t^2+q_{AB}(t)\,\theta^A\theta^B$ with $\theta^A=(\dn T,\sigma^a)$. This is a spatially homogeneous cosmology on $\mathbb R\times SU(2)$, whose evolution is determined entirely by the horizon data. If the BKL conjecture applies to our solutions, we expect that evolution sufficiently deep in the interior will be well-described by a cosmological billiard. The billiards of homogeneous cosmologies of this type (without imposing a horizon) were classified in Ref.~\cite{deBuyl:2003ub}, which we briefly describe.

Near the singularity, we write the spatial metric in Iwasawa form,
\begin{equation}
 q_{AB}\theta^A\theta^B
 =\sum_{A=0}^3 e^{-2\beta^A}
 \bigl(U^A{}_B\theta^B\bigr)^2 \,,
 \label{eq:iwasawa}
\end{equation}
where $U$ is upper triangular with unit diagonal and $\beta^A$ are logarithmic scale factors. With a choice of lapse $V = \sqrt{{\rm det} \, q} = e^{-\sum_A \beta^A}$, the Hamiltonian constraint takes the form~\cite{Damour:2002et}
\begin{equation}
\frac14G^{AB}\pi_A\pi_B+\textstyle\sum_w C_w\,e^{-2w(\beta)}+\dots=0 \,,
\label{eq:hamiltonian}
\end{equation}
where $\pi_A$ are the momenta conjugate to $\beta^A$ and $G^{AB} = \delta^{AB} - \tfrac{1}{3}$ is the inverse of the Lorentzian DeWitt metric. The leading potentials have linear wall forms $w(\beta) = w_A \beta^A$ and coefficients $C_w \ge 0$. The ellipsis denotes terms subdominant near the singularity. Writing $\beta^A = \rho_\beta \gamma^A$, with $\rho_\beta^2 = - G(\beta,\beta)$, separates the approach to the singularity from motion in the hyperbolic space of directions $\gamma^A$. As $\rho_\beta \to \infty$, the dominant exponential potentials become hard walls, confining the dynamics to a billiard. 

In this limit, Eq.~\eqref{eq:hamiltonian} describes a massless particle moving within the chamber bounded by the dominant walls. Between collisions, the motion corresponds to Kasner epochs $\beta^A \simeq p_A x + {\rm const.}$, where $x = \ln (V_{\rm max}/V)$, with $V_{\rm max}$ the maximum interior volume reached before contraction. The definition of $x$ gives $\sum_A p_A = 1$, while the free Hamiltonian constraint gives $\sum_A p_A^2 = 1$. Collisions with the walls produce specular reflections and transitions between Kasner epochs.  

For the $\mathbb{R} \times SU(2)$ cosmology, the possible wall forms $w(\beta)$ are of two types: curvature walls, denoted by $c$, and symmetry walls, denoted by $s$. The intrinsic curvature of the squashed three-spheres gives $c_{abc} = 2 \beta^a + \beta^0$, while the spatial curvature $A_a \dn \sigma^a$ of the Kaluza--Klein connections gives the magnetic walls $c_{0bc} = 2 \beta^0 + \beta^a$. Here $a,b,c$ are distinct elements of $\{1,2,3\}$. Symmetry walls arise from off-diagonal kinetic terms, and come in two types: the electric walls $s_{0a} = \beta^a - \beta^0$ are associated with the momenta of the connection, while $s_{ab} = \beta^b - \beta^a$, with $a < b$, arise from the off-diagonal terms in $h_{ab}$. 

For generic $\mathbb R\times SU(2)$ cosmologies, the dominant walls are those of five-dimensional vacuum gravity~\cite{Damour:2001sa,deBuyl:2003ub},
\begin{equation}
 s_{01},\quad s_{12},\quad s_{23},\quad
 c_{023}=2\beta^0+\beta^1.
 \label{eq:ae4}
\end{equation}
Together they bound the billiard chamber, which is the fundamental Weyl chamber of the hyperbolic Kac–Moody algebra $AE_4$~\cite{Damour:2001sa,Damour:2002et}. This chamber has finite volume, and hence yields chaotic billiard dynamics.  Reflection from a symmetry wall $s_{AB}$ exchanges $p_A$ and $p_B$, while reflection from the curvature wall $c_{023}$ gives
\begin{equation}
\begin{split}
 p'=\frac{1}{1+2p_0+p_1}\bigl(
 &-p_0-p_1,\,p_1,\\
 &p_2+2p_0+p_1,\,p_3+2p_0+p_1
 \bigr) \, .
\end{split}
\label{eq:bkl-map}
\end{equation}
The denominator ensures $\sum_Ap'_A=1$, and the map also
preserves $\sum_A(p'_A)^2=1$.

\begin{table}[t]
\centering\footnotesize
\setlength{\tabcolsep}{2pt}
\begin{tabular}{@{}llll@{}}
\toprule
$A^{\rm H}$ & $H$ & Dominant walls & Billiard\\
\midrule
$(A_1,A_2,A_3)$ & distinct & $c_{023},s_{01},s_{12},s_{23}$ & closed ($AE_4$)\\
$(A_1,A_2,0)$ & $H_1\neq H_2$ & $c_{023},s_{01},s_{12},c_{312}$ & closed (2 cusps)\\
$(A_1,0,0)$ & $H_2\neq H_3$ & $c_{023},s_{01},c_{213},c_{312}$ & open\\
$(A_1,0,0)$ & $H_2=H_3$ & $c_{023},s_{01}$ & open\\
$0$ & distinct & $c_{123},c_{213},c_{312}$ & open\\
$0$ & $H_1\neq H_2=H_3$ & $c_{123}$ & open\\
$0$ & round & none & open\\
\bottomrule
\end{tabular}
\caption{Dominant walls of the interior billiard for each class of horizon data, in the eigenframe of $H_{ab}$. Within degenerate eigenspaces of $H_{ab}$, the frame is rotated to set as many components of $A^{\rm H}$ to zero as possible; the listed components are nonzero. The billiard can be closed only when Eq.~\eqref{eq:criterion} holds.}
\label{tab:billiards}
\end{table}

\begin{figure}[t]
\centering
\includegraphics[width=0.82\columnwidth]{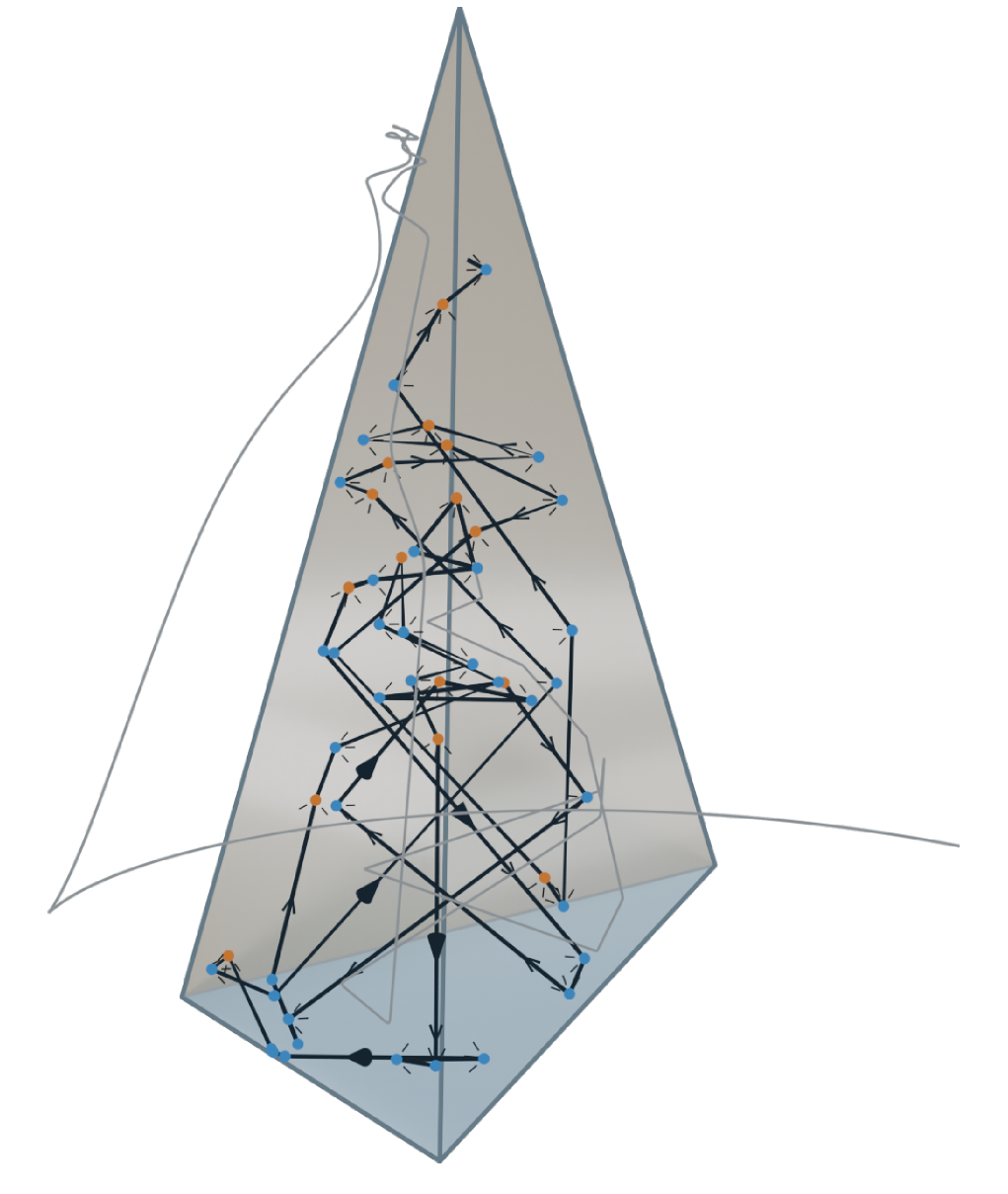}
\caption{The billiard chamber in the hyperbolic space defined by the $\gamma$ directions. The billiard is drawn so that the walls are planes and Kasner epochs are straight line segments. The top vertex is the cusp, where $c_{023}$, $s_{01}$ and $s_{12}$ meet, corresponding to Kasner exponents $(0,0,0, 1)$; the base is $s_{23}$. The curve is the interior of the numerical black hole solution of Fig.~\ref{fig:exterior-solution}, projected onto $\gamma$ with no billiard approximation. Black: the interior solution for $2\times10^3<x<10^8$, with orange and blue dots marking the 52 isolated bounces off curvature and symmetry walls. Grey: the earlier part of the trajectory before the walls become sharp.}
\label{fig:chamber}
\end{figure}

\begin{figure*}[t]
\centering
\includegraphics[width=\textwidth]{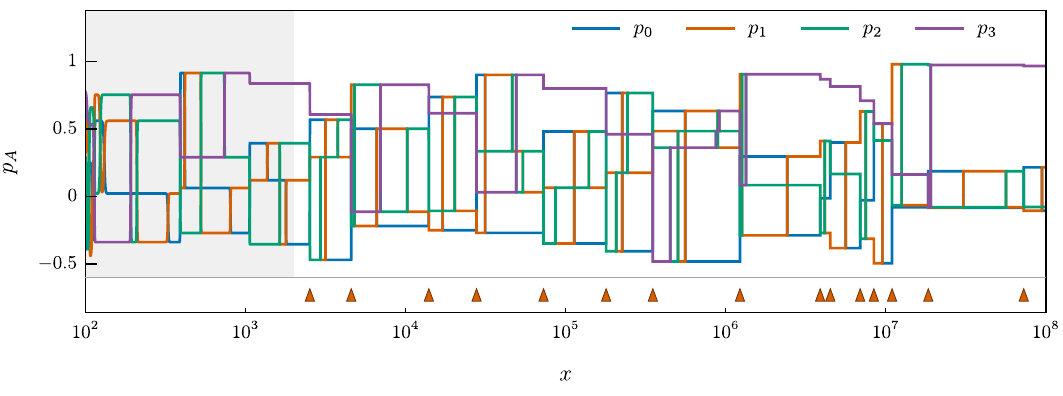}
\caption{Iwasawa exponents $p_A=\dn\beta^A/\dn x$ for the black hole in Fig.~\ref{fig:exterior-solution}, with $x=\ln (V_{\rm max}/V)$. Plateaus are Kasner epochs. Orange triangles in the lower strip indicate reflections off the curvature wall, all other collisions occur with symmetry walls. Beyond the shaded
region ($x<2\times10^3$ where the walls are still soft) there are 52 isolated bounces. }
\label{fig:exponents}
\end{figure*}

Which of walls appear inside the black holes is determined by the horizon data. Since the horizon data determine the full solution, any symmetry of the horizon data is a symmetry of the whole spacetime. Such symmetries can force some fields, and hence walls, to vanish identically. For example, suppose that $A_3^{\rm H} = 0$ in the eigenbasis of $H_{ab}$. Then the transformation $(\sigma^1, \sigma^2, \sigma^3) \to (-\sigma^1, -\sigma^2, \sigma^3)$ combined with $T \to - T$ is a symmetry of the horizon data, and hence of the solution. Away from the horizon, this transformation reverses the signs of $A_3$, $h_{13}$ and $h_{23}$. By symmetry, these must therefore vanish everywhere, removing the walls $c_{012}$, $s_{03}$, $s_{13}$, and~$s_{23}$.

Conversely, nonzero $A_a^{\rm H}$ activates $c_{0bc}$ and $s_{0a}$, and each nonzero pair of components generates the corresponding off-diagonal terms in $h_{ab}$, which behaves like  $h_{ab} \propto H_a H_b A_a^{\rm H} A_b^{\rm H} t^4$ just inside the horizon, activating~$s_{ab}$. In Table~\ref{tab:billiards}  we summarize the dominant walls for each class of horizon data---see also the Appendix.

Realizing the billiard of generic vacuum gravity requires three distinct eigenvalues for the horizon metric $H_{ab}$, with $A_a^{\rm H}$ having nonzero values in each eigenspace. More generally, the interior billiard can be closed only if 
\begin{equation}
A^{\rm H}\times\bigl(HA^{\rm H}\bigr)\neq0 .
\label{eq:criterion}
\end{equation}
This is a necessary condition on the horizon data. We have not proven it to be also sufficient, though exceptions would require fine-tuning.

\emph{Tests of the billiard.}---The billiard and the reflection rule~\eqref{eq:bkl-map} provide predictions that we test against the numerical interior evolution. We integrate the full Einstein equations from the horizon inward. The interior initially expands, reaches a maximum volume, and then begins contraction.   Figures~\ref{fig:chamber} and~\ref{fig:exponents} show the interior of the same black hole as Fig.~\ref{fig:exterior-solution} to $x=10^8$. After an initial regime of overlapping transitions, the evolution resolves into long Kasner epochs separated by isolated bounces. For $2\times10^3<x<10^8$ we identify 52 such bounces: 15 curvature reflections and 37 symmetry reflections, reaching every wall in Eq.~\eqref{eq:ae4}. Between bounces, the Kasner relations hold, the transitions obey Eq.~\eqref{eq:bkl-map}, and we observe no isolated reflections off subdominant walls.


\begin{figure*}[t]
\centering
\includegraphics[width=\textwidth]{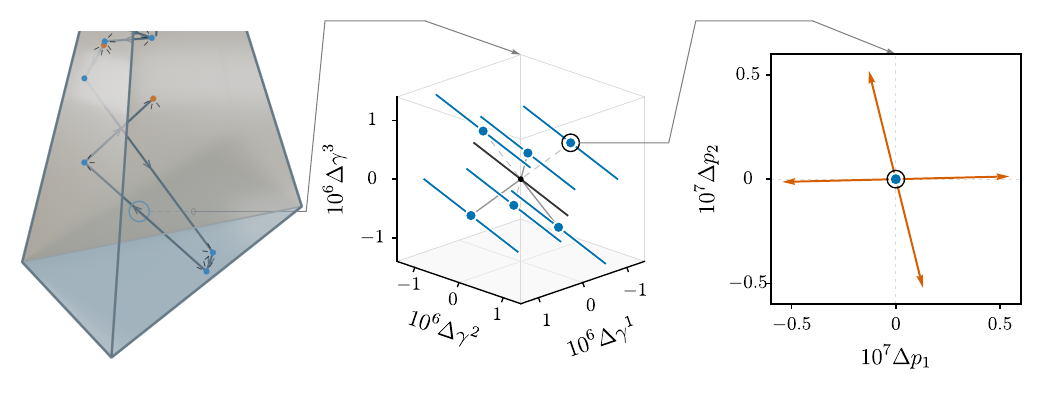}
\caption{Nearby black hole interiors at fixed horizon area.
Left: A partial reproduction of Fig.~\ref{fig:chamber}, highlighting a point along the trajectory.  Middle: nearby trajectories relative to that point; blue dots mark equal $\rho_\beta$, and grey lines join opposite displacements. Right: differences in Kasner exponents $\Delta p_i=p_i-p_i^{\mathrm{blue}}$ ($i=1,2$) for four solutions that intersect the circled blue position at the same depth. Orange arrows represent offsets in the exponents. By tuning horizon data, each of the billiard coordinates can be independently adjusted.}
\label{fig:openness}
\end{figure*}

\emph{The accessible billiard trajectories.}---Regular horizons restrict the initial data of the homogeneous cosmology. Without a horizon, the spatial metric $q_{AB}$ and its derivative $\dot q_{AB}$ supply twenty parameters. The Hamiltonian and momentum constraints reduce this number by four (the momentum constraint along the $\mathbb{R}$ direction vanishes identically). One can further implement three constant $SO(3)$ rotations, set the normalization of the noncompact coordinate $T$, and decide the origin of the time coordinate. In total, this leaves eleven physical parameters in the general case. By contrast, a regular horizon at $t=0$ is characterized by ten coefficients, $(H_{ab},A_a^{\rm H},\kappa)$, reduced to six by rotations and time normalization. Solutions with regular horizons therefore represent a codimension-five subset of the generic local solution space.

This raises the possibility that black holes access only a measure zero set of billiard trajectories. However, the limiting billiard depends on fewer parameters than the general homogeneous solution. This is because the off-diagonal metric variables asymptotically freeze, while the precise values of the wall coefficients $C_w$ do not play a role after the walls become sharp~\cite{Damour:2002et}. At fixed $\rho_\beta$, a trajectory in the billiard is determined by its position, given by $\gamma^A$, and its direction, given by the Kasner exponents $p_A$. The hyperboloid constraint and two Kasner relations leave five independent billiard parameters at fixed $\rho_\beta$. It is therefore possible, in principle, for the black hole solutions to realize the full chaotic billiard. 

To test this, we compare several black hole solutions at fixed depth $\rho_\beta$ and assess whether varying horizon parameters allows us to adjust each of the independent billiard data. We consider black holes with fixed horizon area specified by data $(H_1,H_2,A^{\rm H}_1,A^{\rm H}_2,A^{\rm H}_3)$, with $H_3$ fixed by $\det H$, and evolve each perturbed solution to the same $\rho_\beta$. There, we assess how changing the horizon data adjusts $(\gamma^1, \gamma^2, \gamma^3, p_1, p_2)$, with $\gamma^0$, $p_0$, and $p_3$ fixed by the hyperboloid constraint and the Kasner relations. The response is measured by the $5\times 5$ Jacobian
\begin{equation}
 J=\left.
 \frac{\partial(\gamma^1,\gamma^2,\gamma^3,p_1,p_2)}
 {\partial(H_1,H_2,A^{\rm H}_1,A^{\rm H}_2,A^{\rm H}_3)}
 \right|_{\det H,\,\rho_\beta}.
 \label{eq:openness-jacobian}
\end{equation}
By the inverse function theorem, a nonsingular Jacobian would imply that varying the horizon data produces arbitrary sufficiently small changes in all five billiard coordinates. Evaluating the Jacobian by finite differences near the solution of Eq.~\eqref{eq:example-horizon}, we find $\operatorname{rank}J=5$, supporting that black holes (at least locally) have access to generic billiard trajectories. 

Figure~\ref{fig:openness} illustrates these independent variations. The left panel shows a segment of the billiard trajectory of the solution~\eqref{eq:example-horizon},  highlighting the depth reached at $x = 8500$ ($\rho_\beta = 6779.425$).
The middle panel zooms in near that point on the trajectory and shows six additional numerical solutions constructed such that at the same value of $\rho_\beta$ they arrive at different positions $(\gamma^1, \gamma^2, \gamma^3)$ in the billiard. Finally, the right panel further zooms in along one of these trajectories, and shows a further four trajectories which, at the same value of $\rho_\beta$, probe the two independent Kasner exponent directions. In each case, we have constructed the exterior solutions as well and confirmed they are regular and causal. 

\emph{Discussion.}---We have constructed asymptotically locally AdS$_5$ vacuum black holes with chaotic BKL interiors. The interior of these solutions is governed by the same cosmological billiard as generic five-dimensional vacuum gravity. The five-dimensional setting allows the horizon generator to commute with the $SU(2)$ generators, which avoids the obstruction to obtaining analogous solutions in four dimensions. 

These solutions provide a natural setting for studying BKL singularities inside vacuum black holes. An important question is whether physical conditions such as causality and stability restrict the accessible billiard trajectories, with implications along the lines of~\cite{Hartnoll:2025hly,DeClerck:2025mem}. Another direction would be to construct analogous solutions in different theories. In the vacuum case, this would provide a clean setting to explore the role of higher-curvature corrections to the BKL picture inside black holes~\cite{Bueno:2024fzg, Caceres:2024edr,Bueno:2024qhh}.  

More broadly, it would be worthwhile to understand which higher-dimensional homogeneous cosmologies permit black hole horizons, and how or if the existence of that horizon constrains the interior chaos. We have not yet found asymptotically AdS solutions with round conformal boundaries, or asymptotically flat solutions, with the same chaotic interiors. It would be interesting to construct those solutions or identify a rigorous obstruction.

Finally, because our solutions are asymptotically locally AdS$_5$, their chaotic interiors should be encoded in thermal states of a holographic CFT on the squashed and twisted boundary geometry~\eqref{eq:boundary}. Spacelike singularities are known to leave signatures in analytically continued thermal correlators~\cite{Fidkowski:2003nf,Festuccia:2005pi,Frenkel:2020ysx}, high-overtone quasinormal modes~\cite{Hartnoll:2026vhu,Xiao:2026pir,Coviello:2026tol,Coviello:2026hdc}, and thermal OPE data~\cite{Ceplak:2024bja,Ceplak:2026ard}. Our solutions provide a vacuum setting in which to determine whether these---or other---probes can resolve the chaotic sequence of Kasner epochs.

\begin{acknowledgments}
\emph{Acknowledgments.}---
We thank Sean Hartnoll and Simon Ross for reading an earlier draft of this manuscript and providing feedback. The work of RAH was supported by a Willmore Fellowship at Durham University and by an STFC Ernest Rutherford Fellowship UKRI5636. AKP is supported by the European Union's Horizon Europe research and innovation programme under the Marie Sk\l{}odowska-Curie grant agreement No.~101210745.

We used OpenAI's GPT-6 Astra and Anthropic's Claude Opus 5.5 to assist with numerical code development and debugging, figure preparation, checking calculations, and editing text. The authors directed these tasks, reviewed the resulting material,  and take responsibility for the contents of this paper.

\end{acknowledgments}

\clearpage
\onecolumngrid
\begin{center}
  \Large\bfseries Appendices
\end{center}
\appendix
\setcounter{secnumdepth}{1}

\begingroup
\makeatletter
\let\appendixtocsection\l@section
\let\l@section\@gobbletwo
\let\l@title\@gobbletwo
\let\l@abstract\@gobbletwo
\let\l@subsection\@gobbletwo
\let\l@subsubsection\@gobbletwo
\renewcommand{\tocname}{Contents}
\tableofcontents
\makeatother
\endgroup
\addtocontents{toc}{\protect\startappendixtoc}

\section{$SU(2)$ conventions}
\label{app:su2_conventions}

We parametrize the elements $g \in SU(2)$ by Euler angles,
\begin{equation}
g = e^{i \phi \, \tau_3/2} e^{i \theta \, \tau_2/2} e^{i \psi \, \tau_3/2} \,, \qquad \theta \in [0, \pi] \,, \quad \phi \in [0, 2 \pi) \,, \quad \psi \in [0, 4 \pi) \, 
\end{equation}
with $\tau_a$ the Pauli matrices. The left-invariant forms are defined by $g^{-1} \dn g = (i/2) \sigma^a \tau_a$, which gives
\begin{align}
\sigma^1 = -\sin\psi\,\dn\theta+\cos\psi\sin\theta\,\dn\phi \, , \qquad 
\sigma^2 = \cos\psi\,\dn\theta+\sin\psi\sin\theta\,\dn\phi \, ,\qquad 
\sigma^3 = \dn\psi+\cos\theta\,\dn\phi \, .
\end{align}
The dual frame, $\sigma^a(e_b)=\delta^a{}_b$, is
\begin{equation}
\begin{aligned}
e_1 = -\sin\psi\,\partial_\theta+\frac{\cos\psi}{\sin\theta}\,\partial_\phi-\cos\psi\cot\theta\,\partial_\psi \, ,\qquad
e_2 =  \cos\psi\,\partial_\theta+\frac{\sin\psi}{\sin\theta}\,\partial_\phi-\sin\psi\cot\theta\,\partial_\psi \, ,\qquad
e_3 = \partial_\psi \, ,
\end{aligned}
\end{equation}
and $e_a$ generates right multiplication, $e_a(g)=g\,\tfrac i2\tau_a$. Left multiplication is generated by
\begin{equation}
\begin{aligned}
\xi_1=-\sin\phi\,\partial_\theta-\cos\phi\cot\theta\,\partial_\phi+\frac{\cos\phi}{\sin\theta}\,\partial_\psi\,,\qquad
\xi_2=\cos\phi\,\partial_\theta-\sin\phi\cot\theta\,\partial_\phi+\frac{\sin\phi}{\sin\theta}\,\partial_\psi\,,\qquad
\xi_3=\partial_\phi\,,
\end{aligned}
\label{eq:su-killing-vecs}
\end{equation}
which act as $\xi_1(g)=-\tfrac i2\tau_1g$ and $\xi_{2,3}(g)=\tfrac i2\tau_{2,3}g$.  In these coordinates $\tfrac14\delta_{ab}\sigma^a\sigma^b=\tfrac14\bigl(\dn\theta^2+\dn\phi^2+\dn\psi^2+2\cos\theta\,\dn\phi\,\dn\psi\bigr)$ is the round metric of the unit sphere.

Since $\mathcal L_{\xi_a}\sigma^b=0$, the $\xi_a$ generate an $SU(2)$ isometry of every metric in our ansatz, and they commute with the $e_a$, $[e_a,\xi_b]=0$. The $e_a$ themselves are not Killing in general, because their flows rotate the coframe. For example,
\begin{equation}
 \mathcal L_{e_3}\sigma^1=-\sigma^2,\qquad
 \mathcal L_{e_3}\sigma^2=\sigma^1,\qquad
 \mathcal L_{e_3}\sigma^3=0,
 \label{eq:e3-rotation}
\end{equation}
so $e_3=\partial_\psi$ is Killing precisely when $h_{11}=h_{22}$, $h_{12}=h_{13}=h_{23}=0$ and $A_1=A_2=0$. It then generates an additional $U(1)$ isometry commuting with $SU(2)$.

\section{Field equations}
The metric ansatz in its Kaluza-Klein form reads
\begin{equation}
 \dn s^2=\dn\rho^2-F^2(\dn T+A_a\sigma^a)^2+h_{ab}\sigma^a\sigma^b \, .
 \label{eq:metric}
\end{equation}
Here $h$ is positive definite and $F,A_a,h_{ab}$ depend only on $\rho$. A horizon with metric $H$ has area $16\pi^2\sqrt{\det H}$.  The field equations are most conveniently expressed in matrix form. Set $\theta^A=(\dn T,\sigma^a)$ and write $\dn s^2=\varepsilon\dn u^2+q_{AB}(u)\theta^A\theta^B$, where $\varepsilon=+1$, $u=\rho$ in the exterior, and $\varepsilon=-1$, $u=t$ in the interior. In the exterior,
\begin{equation}
 q_{00}=-F^2,\qquad q_{0a}=-F^2A_a,\qquad q_{ab}=h_{ab}-F^2A_aA_b \, .
\end{equation}
Define $K^A{}_B=\tfrac12q^{AC}\dn q_{CB}/\dn u$ and $K=K^A{}_A$. If $\mathcal R_{AB}$ is the Ricci tensor of the homogeneous orbit metric $q$, the evolution equations and constraints are
\begin{align}
 \frac{\dn q_{AB}}{\dn u}&=2q_{AC}K^C{}_B,&
 \frac{\dn K^A{}_B}{\dn u}&=\varepsilon\bigl(\mathcal R^A{}_B+4L^{-2}\delta^A{}_B\bigr)-KK^A{}_B,
 \label{eq:evolution}\\
 0&=K^2-K^A{}_BK^B{}_A-\varepsilon(\mathcal R+12L^{-2}),&
 0&=C^C{}_{AB}K^B{}_C \, .
 \label{eq:constraints}
\end{align}
The structure constants are defined by $\dn\theta^A=-\tfrac12C^A{}_{BC}\theta^B\wedge\theta^C$, so $C^a{}_{bc}=-\epsilon_{abc}$ and every component involving the index $0$ vanishes. The curvature is algebraic in the metric and the structure constants and, for either signature of $q$,
\begin{equation}
 \mathcal R_{AB}=-\frac12C^C{}_{AD}C^D{}_{BC}
 -\frac12q^{CD}q_{EF}C^E{}_{AC}C^F{}_{BD}
 +\frac14q^{CE}q^{DF}q_{AG}q_{BH}C^G{}_{CD}C^H{}_{EF} \, .
 \label{eq:orbit-ricci}
\end{equation}
Equations~\eqref{eq:evolution}--\eqref{eq:orbit-ricci} specify the full vacuum Einstein equations. There are three potentially nontrivial momentum constraints: the $A=0$ equation vanishes identically.

\section{Near-horizon solution}

A smooth nonextremal horizon at $\rho=0$ requires $F=\kappa\rho+O(\rho^3)$, with $F/\rho$, $A_a$ and $h_{ab}$ smooth functions of $\rho^2$. To understand this requirement, introduce local Rindler coordinates $X=\rho\cosh(\kappa T)$ and $Y=\rho\sinh(\kappa T)$, in terms of which the leading $(\rho,T)$ part of the metric becomes $\dn X^2-\dn Y^2$, while
\begin{equation}
    \rho^2=X^2-Y^2,\qquad
 \kappa\rho^2\dn T=X\dn Y-Y\dn X \, .
\end{equation}
Since $\rho$ itself has a square root dependence on the Kruskal coordinates, a smooth extension therefore requires $\rho$ to appear in the metric components in even powers. Examining this local Kruskal extension also explains why the interior expansion follows by continuing $\rho^2 \to -t^2$. 

Solving the field equations in the vicinity of the horizon can be carried out straightforwardly. We begin by choosing a frame in which the horizon metric is diagonal $H_{ab} = {\rm diag}(H_1, H_2, H_3)$, and denote the connection in this frame as $A_a^{\rm H}$. Let $\mathcal R_{ab}[H]$ and $\mathcal R[H]$ denote the Ricci tensor and scalar of $H_{ab}\sigma^a\sigma^b$. In the frame in which $H_{ab}$ is diagonal, so too is its curvature, which has non-vanishing components
\begin{equation}
 \mathcal R_{aa}[H]
 =\frac{H_a^2-(H_b-H_c)^2}{2H_bH_c},
 \qquad \{a,b,c\}=\{1,2,3\},
\end{equation}
with no sum over $a$.  Then the leading near horizon solution of the equations of motion takes the form
\begin{align}
 F&=F_{(1)}\rho+F_{(3)}\rho^3+F_{(5)}\rho^5+O(\rho^7),\nonumber\\
 h_{ab}&=h_{ab}^{(0)}+h_{ab}^{(2)}\rho^2+h_{ab}^{(4)}\rho^4+O(\rho^6),\nonumber\\
 A_a&=A_a^{(0)}+A_a^{(2)}\rho^2+A_a^{(4)}\rho^4+O(\rho^6).
 \label{eq:horizon}
\end{align}
We include here explicitly the fourth order contributions because they will have a role in the analysis of the billiard walls below. For the lapse the coefficients read
\begin{align}
 F_{(1)}={}&\kappa,\qquad
 F_{(3)}=-\kappa\left(\frac{1}{3L^2}+\frac{\mathcal R[H]}{12}\right),\nonumber\\
 F_{(5)}={}&\kappa\left[\frac{19}{30L^4}
 +\frac{\mathcal R[H]}{6L^2}+\frac{\mathcal R[H]^2}{120}
 +\frac{1}{160}\sum_{i=1}^3\frac{\mathcal R_{ii}[H]^2}{H_i^2}
 -\frac{3\kappa^2}{80\det H}\sum_{i=1}^3H_i(A_i^{\rm H})^2\right] \,. 
 \label{eq:horizon-lapse-coefficients}
\end{align}
For the angular metric, the first two coefficients are
\begin{equation}
 h_{ab}^{(0)}=H_{ab},\qquad
 h_{ab}^{(2)}=\frac12\mathcal R_{ab}[H]+\frac{2}{L^2}H_{ab} \, .
 \label{eq:horizon-h2}
\end{equation}
To this order, $h_{ab}$ remains diagonal. This changes at fourth order, and it is easiest to separate the contributions to the diagonal and off-diagonal terms. The diagonal coefficient at forth order is
\begin{align}
 h_{aa}^{(4)}={}&\frac{2H_a}{3L^4}
 +\frac{5\mathcal R_{aa}[H]-H_a\mathcal R[H]}{12L^2}
 +\frac{\mathcal R_{aa}[H]^2}{8H_a}
 -\frac{\mathcal R[H]\mathcal R_{aa}[H]}{12} +\frac{H_a\left[H_a\mathcal R_{aa}[H]
 -(H_b-H_c)(\mathcal R_{bb}[H]-\mathcal R_{cc}[H])\right]}{16\det H}
 \nonumber 
 \\
 &+\frac{\kappa^2H_a\left[H_b(A_b^{\rm H})^2+H_c(A_c^{\rm H})^2\right]}{16\det H} 
 \label{eq:horizon-h-coefficients}
\end{align}
where $\{a,b,c\} = \{1,2,3\}$ in that order, with cyclic permutations giving the other two diagonal componentss. No sum is implied over repeated indices. The off-diagonal contribution, for $a \neq b$, is
\begin{equation}
 h_{ab}^{(2)}=0,\qquad
 h_{ab}^{(4)}=-\frac{\kappa^2H_aH_bA_a^{\rm H}A_b^{\rm H}}{16\det H}.
 \label{eq:horizon-off-diagonal}
\end{equation}
Finally, the connection coefficients are
\begin{align}
 A_a^{(0)}={}&A_a^{\rm H},\qquad
 A_a^{(2)}=\frac{H_a^2A_a^{\rm H}}{8\det H},\nonumber\\
 A_a^{(4)}={}&\frac{H_a^2A_a^{\rm H}}{192\det H}
 \left[\frac{H_a^2}{\det H}-\frac{16}{L^2}-4\mathcal R[H]
 +\frac{10\mathcal R_{aa}[H]}{H_a}\right].
 \label{eq:horizon-a-coefficients}
\end{align}

The near-horizon solution can be extended to higher orders by  substituting an even power series for $\rho$ into Eq.~\eqref{eq:evolution}. The solution is fully determined by $(\kappa, H_{ab}, A_a^{\rm H})$, and no further integration constants occur. The freedom to choose the normalization of the time coordinate can be used to set $\kappa = 1$. Doing so results in a boundary time coordinate that is not canonically normalized, but this can be easily adjusted after the full numerical solution has been obtained. 

The interior solution is initialized separately by continuing $\rho^2\to-t^2$ in this series. Thus $q_{00}=\kappa^2t^2+O(t^4)$ and the interior spatial metric is positive definite just inside the horizon. We test the consequences of using different choices of initialization time $t$ and truncating the near-horizon solution at different orders in our numerical implementation. 

\section{A sufficient condition to exclude inner Killing horizons}

As our solutions are numerically constructed, the interior evolution never actually \textit{reaches} the singularity. This leaves open the possibility that, rather than terminating at a singularity, the evolution  may reach  an inner Cauchy horizon instead. Here we derive a sufficient condition that excludes non-degenerate inner Killing horizons. Throughout this appendix, a dot denotes proper time differentiation, $K=K^A{}_A=\dot V/V$, and $q$ is positive definite over the interior.

In Eq.~\eqref{eq:orbit-ricci}, the first two contributions to $\mathcal R_{A0}$ vanish because the index $0$ enters a structure constant. Raising the first index to obtain $\mathcal R^0{}_0$ makes the final contribution proportional to $C^0{}_{BC}$, which also vanishes. Thus $\mathcal R^0{}_0=0$, although $\mathcal R_{00}$ need not vanish. Equation~\eqref{eq:evolution} then implies
\begin{equation}
 Q\equiv VK^0{}_0,\qquad \dot Q=-\frac{4V}{L^2} \, .
 \label{eq:central-current-evolution}
\end{equation}
In terms of the metric variables, the quantity $K^0{}_0$ reads 
\begin{equation}
 K^0{}_0=\frac12\frac{\dot q_{00}}{q_{00}}
 -\frac12q_{00}A^Th^{-1}\dot A \, .
 \label{eq:central-current-metric}
\end{equation}
At the outer horizon, $q_{00}=\kappa^2t^2+O(t^4)$ and $V=\kappa t\sqrt{\det H}+O(t^3)$, so $Q\to\kappa\sqrt{\det H}>0$. At a smooth inner horizon of $\partial_T$, introduce the time $\tau=t_--t$. Regularity gives there $q_{00}=\kappa_-^2\tau^2+O(\tau^4)$ and $V=\kappa_-\tau\sqrt{\det H^-}+O(\tau^3)$, with $\kappa_->0$ and $H^-$ positive definite. Since $\dot\tau=-1$, Eq.~\eqref{eq:central-current-metric} then requires $Q\to-\kappa_-\sqrt{\det H^-}<0$ at such an inner horizon. In both limits $A$ and $h$ are smooth as functions of $t^2$ or $\tau^2$, and the second term in Eq.~\eqref{eq:central-current-metric} does not contribute. An inner horizon of $\partial_T$ therefore requires a sign change of $Q$. 

Such a sign change is not automatically ruled out. Because the right-hand side of Eq.~\eqref{eq:central-current-evolution} is negative and non-zero, $Q$ can actually decrease. We can, however, bound the amount by which $Q$ can decrease. The trace of Eq.~\eqref{eq:evolution}, combined with the Hamiltonian constraint, gives
\begin{equation}
 \dot K=-K^A{}_BK^B{}_A-\frac4{L^2}\le-\frac{K^2}{4} \, .
 \label{eq:inner-raychaudhuri}
\end{equation}
The inequality follows because, in an orthonormal frame for the positive-definite metric $q$, $K^A{}_B$ becomes a real symmetric matrix. Its four eigenvalues are therefore real, and by the Cauchy--Schwarz inequality the sum of their squares is at least $K^2/4$. Suppose $K(t_0)<0$ and set $s_0=|K(t_0)|$. Integrating Eq.~\eqref{eq:inner-raychaudhuri} and $\dot V=KV$ gives
\begin{equation}
 V(t)\le V(t_0)\left[1-\frac{s_0}{4}(t-t_0)\right]^4.
 \label{eq:inner-volume-bound}
\end{equation}
Thus, the volume must reach zero at or before $t =  t_0+4/s_0$. This then allows us to bound the change in $Q$. Assuming the positive upper bound in Eq.~\eqref{eq:inner-volume-bound}  and integrating Eq.~\eqref{eq:central-current-evolution}
\begin{equation}
 Q(t)\ge Q(t_0)-\frac{16V(t_0)}{5L^2|K(t_0)|} \, .
\end{equation}
It follows that if a contracting timeslice satisfies
\begin{equation}
Q(t_0)>\frac{16V(t_0)}{5L^2|K(t_0)|}
 \label{eq:inner-criterion}
\end{equation}
then $Q$ can never go negative, which
excludes a smooth, nondegenerate inner horizon generated by $\partial_T$.

One may worry that this condition does not exclude inner Killing horizons with a different generator. Any other generator commuting with $\mathbb R\times SU(2)$ would have to include a rotation of the sphere, which is a symmetry only if the horizon data are invariant under that rotation. This requires $A^{\rm H}$ to be an eigenvector of $H$, and so is excluded by Eq.~\eqref{eq:criterion}. The equal-spinning Myers--Perry solution shows that this restriction is necessary. Its connection lies along the symmetry axis of its biaxial horizon, so Eq.~\eqref{eq:criterion} fails and rotations about that axis, generated by $e_3$, are an additional symmetry. Its two horizons rotate at different angular velocities: with $T$ normalized so that $\partial_T$ generates the outer horizon, the inner horizon is generated by $\partial_T+\Omega e_3$ with $\Omega\neq0$, so $\partial_T$ is not null there and $Q$ need not change sign. For horizon data satisfying Eq.~\eqref{eq:criterion}, no such additional symmetry exists, and the criterion~\eqref{eq:inner-criterion} therefore excludes every smooth nondegenerate inner Killing horizon whose generator commutes with $\mathbb R\times SU(2)$.

For the horizon data~\eqref{eq:example-horizon}, at $x=1$ in the numerical clock of Appendix~\ref{app:numerics},
\begin{equation}
 Q(t_0)=1.51618517383,\qquad
 \frac{16V(t_0)}{5L^2|K(t_0)|}=0.07945366039,
 \label{eq:inner-example}
\end{equation}
which satisfied the sufficient condition by a margin of $1.4367315134$. Thus, we can be assured that that spacetime does not end at an inner Killing horizon, and the evolution is consistent with a spacelike singularity.

It is worth remarking on how the result here compares with existing no-inner-horizon theorems in the literature. Those theorems, for black holes with scalar hair, rest on a quantity whose values at an outer and an inner horizon must have opposite signs, and on a matter field that forces both horizons to share a generator. In the static charged-scalar case, the scalar potential cancels from the relevant identity, so planar and spherical inner horizons are excluded even in AdS~\cite{Hartnoll:2020fhc,Cai:2020wrp,An:2021plu}. The argument of Ref.~\cite{Dias:2021afz} for rotating black holes instead integrates the Killing identity for the common generator between the two horizons, and obtains the contradiction from a positive scalar potential. Our argument has the same structure, with Eq.~\eqref{eq:criterion} fixing the generator within the symmetry class considered, but it lacks the sign contradiction that makes those results robust.  Here, the negative cosmological constant gives $\dot Q=-4V/L^2<0$, which permits the sign change required at an inner horizon. Excluding inner horizons therefore requires the additional bound~\eqref{eq:inner-criterion}, obtained from the Raychaudhuri equation. This bound must be checked for each solution. 

It may be the case that a general proof can be found. In fact, in our numerics we have purposefully constructed solutions trying to violate the bound~\eqref{eq:inner-criterion}. Such solutions exist, but all of those we have constructed (within the class of horizon data satisfying~\eqref{eq:criterion}) have closed timelike curves, satisfying $v^2 > 1$ somewhere in the exterior, and hence are physically excluded. However, even in these cases when we evolve the solutions into the interior we find growing curvature and no signs of an inner horizon. This could be that our search has not yet been thorough enough, or a hint that a stronger condition than~\eqref{eq:inner-criterion} exists.

\section{Billiard walls and the horizon data}

In this section we explain in more detail the walls that bound the cosmological billiard and how exact symmetries of the solution can be used to explicitly exclude certain walls altogether. We refer to~\cite{Damour:2002et, Belinski:2017fas} for more details on cosmological billiards in general. 

A wall can be absent from the billiard for three reasons:
(i) its coefficient vanishes identically;
(ii) an exact symmetry makes it cancel against another term with the same exponential; or
(iii) it is redundant, meaning that its inequality is implied by other walls that are present. The horizon data control (i) and (ii), through discrete and continuous symmetries of the solution. Item (iii) follows from linear algebra on the set of all wall forms. We begin by presenting the potential and walls, and then discuss the three mechanisms in turn.

\subsection{The Iwasawa decomposition} 
We express components in the fixed basis $\theta^A=(\dn T,\sigma^a)$, chosen at the horizon so that $H_{ab}$ is diagonal. When $H_{ab}$ has a repeated eigenvalue, this leaves a residual freedom, which we fix below Eq.~\eqref{eq:adapted-horizon}. The basis is not rotated during the evolution, so $h_{ab}(t)$ need not remain diagonal. We write the interior spatial metric in the Iwasawa form~\eqref{eq:iwasawa}, $q=U^TDU$ with $D=\operatorname{diag}(e^{-2\beta^0},\dots,e^{-2\beta^3})$. For our ansatz the unit upper-triangular matrix $U$ has the block form
\begin{equation}
 U=\begin{pmatrix}1&A^T\\0&N\end{pmatrix},
 \qquad
 N=\begin{pmatrix}1&n_{12}&n_{13}\\0&1&n_{23}\\0&0&1\end{pmatrix},
 \qquad
 h=N^T\operatorname{diag}\bigl(e^{-2\beta^1},e^{-2\beta^2},e^{-2\beta^3}\bigr)N ,
 \label{eq:angular-iwasawa}
\end{equation}
so that the interior metric reads
$-\dn t^2+e^{-2\beta^0}(\dn T+A_a\sigma^a)^2+h_{ab}\sigma^a\sigma^b$.
The first row of $U$ is therefore the Kaluza--Klein connection itself, and $N$ is the Iwasawa factor of the angular metric. Explicitly,
\begin{equation}
 n_{12}=\frac{h_{12}}{h_{11}},\qquad
 n_{13}=\frac{h_{13}}{h_{11}},\qquad
 n_{23}=\frac{h_{23}-h_{12}h_{13}/h_{11}}{h_{22}-h_{12}^2/h_{11}},
 \label{eq:angular-mixing}
\end{equation}
and the logarithmic scale factors are
\begin{equation}
 e^{-2\beta^0}=q_{00},\qquad e^{-2\beta^1}=h_{11},\qquad
 e^{-2\beta^2}=h_{22}-\frac{h_{12}^2}{h_{11}},\qquad
 e^{-2\beta^3}=\frac{\det h}{h_{11}h_{22}-h_{12}^2}.
 \label{eq:metric-scale-factors}
\end{equation}
The Iwasawa variables depend on how we order the basis vectors. Our convention is to label the basis so that the nonzero components of the horizon connection come first---this convention is discussed below Eq.~\eqref{eq:adapted-horizon}. Throughout this section, a dot denotes differentiation with respect to the interior proper time $t$, and $V=\sqrt{\det q}=\sqrt{q_{00}\det h}=e^{-\sum_A\beta^A}$ is the volume density. The Hamiltonian is written using the clock $\eta$, defined by $\dn t=V\dn\eta$.

\subsection{The Hamiltonian constraint}
In the interior ($\varepsilon=-1$), the Hamiltonian constraint in Eq.~\eqref{eq:constraints} reads $0=K^2-K^A{}_BK^B{}_A+\mathcal R+12L^{-2}$. To express its kinetic part in Iwasawa variables, substitute $q=U^TDU$ into $K^A{}_B=\tfrac12(q^{-1}\dot q)^A{}_B$. This gives
\begin{equation}
 q^{-1}\dot q=U^{-1}\bigl(\dot DD^{-1}+\Omega+D^{-1}\Omega^TD\bigr)U,
 \qquad \text{where} \qquad \Omega\equiv\dot UU^{-1}.
\end{equation}
Here $\dot DD^{-1}=-2\operatorname{diag}(\dot\beta^A)$ is diagonal, $\Omega$ is strictly upper triangular, and $D^{-1}\Omega^TD$ is strictly lower triangular. Taking traces then gives
\begin{equation}
 V^2\bigl(K^A{}_BK^B{}_A-K^2\bigr)
 =V^2G_{AB}\dot\beta^A\dot\beta^B
 +\frac{V^2}{2}\sum_{A<B}e^{2(\beta^B-\beta^A)}\,\Omega_{AB}^2,
 \qquad G_{AB}=\delta_{AB}-1 .
 \label{eq:kinetic-split}
\end{equation}
We then multiply the constraint by $-V^2$ and define $\pi_A=2G_{AB}\,\dn\beta^B/\dn\eta$, so that $V^2G_{AB}\dot\beta^A\dot\beta^B=\tfrac14G^{AB}\pi_A\pi_B$ with $G^{AB}=\delta^{AB}-\tfrac13$. The result is
\begin{equation}
 0=\frac14G^{AB}\pi_A\pi_B
 +\sum_{A<B}C_{s_{AB}}e^{-2s_{AB}(\beta)}
 -V^2\mathcal R-\frac{12}{L^2}V^2,
 \label{eq:full-hamiltonian}
\end{equation}
where
\begin{equation}
 s_{AB}=\beta^B-\beta^A,
 \qquad C_{s_{AB}}=\frac12\mathcal J_{AB}^2,
 \qquad \mathcal J_{AB}=Ve^{2(\beta^B-\beta^A)}\Omega_{AB} .
 \label{eq:symmetry-coefficients}
\end{equation}
Here $\mathcal R$ is the scalar curvature of the four-dimensional spatial metric $q$. Equation~\eqref{eq:full-hamiltonian} is the complete Hamiltonian constraint. When the potential terms are negligible, the motion in $\beta$-space is free and null, $G^{AB}\pi_A\pi_B=0$, and corresponds to a Kasner epoch. We now identify each potential term.

\begin{table}[t]
\centering\small
\renewcommand{\arraystretch}{1.3}
\begin{tabular}{@{}llll@{}}\toprule
Type & Wall & Form $w(\beta)$ & Coefficient $C_w$\\\midrule
Magnetic & $c_{023}$ & $2\beta^0+\beta^1$ & $\frac12\bigl(A_1+n_{12}A_2+n_{13}A_3\bigr)^2$\\
 & $c_{013}$ & $2\beta^0+\beta^2$ & $\frac12\bigl(A_2+n_{23}A_3\bigr)^2$\\
 & $c_{012}$ & $2\beta^0+\beta^3$ & $\frac12A_3^2$\\\midrule
Sphere & $c_{123}$ & $\beta^0+2\beta^1$ & $\frac12\bigl(1+n_{12}^2+n_{13}^2\bigr)^2$\\
 & $c_{213}$ & $\beta^0+2\beta^2$ & $\frac12\bigl(1+n_{23}^2\bigr)^2$\\
 & $c_{312}$ & $\beta^0+2\beta^3$ & $\frac12$\\\midrule
Electric & $s_{01}$ & $\beta^1-\beta^0$ & $\frac{V^2}{2}e^{4w}\,\bigl(\dot A_1\bigr)^2$\\
 & $s_{02}$ & $\beta^2-\beta^0$ & $\frac{V^2}{2}e^{4w}\,\bigl(\dot A_2-n_{12}\dot A_1\bigr)^2$\\
 & $s_{03}$ & $\beta^3-\beta^0$ & $\frac{V^2}{2}e^{4w}\,\bigl(\dot A_3-n_{23}\dot A_2+(n_{12}n_{23}-n_{13})\dot A_1\bigr)^2$\\\midrule
Angular & $s_{12}$ & $\beta^2-\beta^1$ & $\frac{V^2}{2}e^{4w}\,\bigl(\dot n_{12}\bigr)^2$\\
 & $s_{13}$ & $\beta^3-\beta^1$ & $\frac{V^2}{2}e^{4w}\,\bigl(\dot n_{13}-n_{23}\dot n_{12}\bigr)^2$\\
 & $s_{23}$ & $\beta^3-\beta^2$ & $\frac{V^2}{2}e^{4w}\,\bigl(\dot n_{23}\bigr)^2$\\\bottomrule
\end{tabular}
\caption{The twelve walls. Each row contributes $C_we^{-2w(\beta)}$ to the Hamiltonian~\eqref{eq:full-hamiltonian}. For the symmetry walls, $e^{4w}$ denotes $e^{4w(\beta)}$, and the bracket is $\Omega_{AB}=(\dot UU^{-1})_{AB}$. Every coefficient is a positive factor times a perfect square, so on a regular interval $C_w$ vanishes identically if and only if the squared quantity does. The sphere coefficients never vanish. Together with the cross terms~\eqref{eq:curvature-cross-terms}, the first six rows give the complete curvature potential.}
\label{tab:wall-catalogue}
\end{table}

\subsection{Symmetry walls}
The second term in Eq.~\eqref{eq:full-hamiltonian} is the kinetic energy of the off-diagonal variables $A_a$ and $n_{ab}$; the relevant entries of $\Omega=\dot UU^{-1}$ are listed in Table~\ref{tab:wall-catalogue}. Following Ref.~\cite{Damour:2002et}, we treat $C_{s_{AB}}=\tfrac12\mathcal J_{AB}^2$ as the wall coefficient and $s_{AB}$ as the wall form, because in canonical variables $\mathcal J_{AB}$ depends only on the off-diagonal variables and their conjugate momenta, with no explicit dependence on $\beta$. For example, $\dot n_{13}$ appears only in $\Omega_{13}$, so $\mathcal J_{13}$ is the momentum conjugate to $n_{13}$,  and similarly for $\mathcal J_{03}$ and $A_3$, and for $\mathcal J_{23}$ and $n_{23}$. These momenta tend to constants as the singularity is approached, whereas the velocities decay. Inverting Eq.~\eqref{eq:symmetry-coefficients} gives $\Omega_{AB}=e^{-2s_{AB}(\beta)}\mathcal J_{AB}/V$, so deep inside the chamber $A$ and $N$ freeze while $C_{s_{AB}}$ stays finite. 

Whether a given symmetry wall is present is therefore a statement about $\Omega_{AB}$. Throughout the interior, between the horizon and the singularity, $q$ is nondegenerate, so $V>0$ and the prefactor $Ve^{2s_{AB}(\beta)}$ in $\mathcal J_{AB}$ is strictly positive. It vanishes only at the endpoints: at the horizon, where $V\simeq t\sqrt{\det H}$, and at the singularity, where $V\to0$. Hence, in the interior, $C_{s_{AB}}$ vanishes at a given time exactly when $\Omega_{AB}$ does, and vanishes identically only when $\Omega_{AB}$ vanishes identically. By Table~\ref{tab:wall-catalogue}, each $\Omega_{AB}$ is an explicit combination of $\dot A_a$ and $\dot n_{ab}$, so deciding whether a symmetry wall is present amounts to whether that combination vanishes for all $t$. This is addressed by the horizon symmetries below. 

There are two kinds of symmetry walls, and they have simple interpretations.  The electric walls $s_{0a}$ come from the first row of $U$, which is the connection, so $\Omega_{0a}$ involves only $\dot A_a$. The momentum conjugate to $A_a$ is $E_a=Ve^{-2\beta^0}h^{ab}\dot A_b$, which is the electric field of the Kaluza--Klein gauge field $A_a\sigma^a$, and one finds $\mathcal J_{0a}=(NE)_a$; that is, $\mathcal J_{0a}$ is this electric field expressed in the Iwasawa frame. An electric wall is therefore present precisely when the connection evolves. The angular walls $s_{ab}$ come from $N$, and $\Omega_{ab}$ involves only $\dot n_{ab}$. At an instant where $h$ is diagonal, meaning $N=I$, $\mathcal J_{ab}=V\dot h_{ab}/h_{bb}$. Hence, the angular walls are present only when the off-diagonal terms in $h_{ab}$ are evolving. 

\subsection{Curvature walls} The spatial curvature can be easily evaluated from the Kaluza-Klein form of the spatial metric $q$. Define the field strength $\mathcal F=\dn(A_a\sigma^a)=\tfrac12A_a\epsilon_{abc}\sigma^b\wedge\sigma^c$, then the four-dimensional spatial curvature takes the form $\mathcal R=\mathcal R[h]-\tfrac14e^{-2\beta^0}\mathcal F_{ab}\mathcal F^{ab}$, with indices raised by $h^{ab}$. The first term, $\mathcal R[h]$, is just the spatial curvature of the Bianchi IX space, $\det h\,\mathcal R[h]=\tfrac12\bigl[(\operatorname{tr}h)^2-2\operatorname{tr}(h^2)\bigr]$. The second term is just $\mathcal F_{ab}\mathcal F^{ab}=2A^ThA/\det h$. Using $V^2=e^{-2\beta^0}\det h$, these give
\begin{equation}
 -V^2\mathcal R
 =\frac{e^{-2\beta^0}}{2}\left[2\operatorname{tr}(h^2)-(\operatorname{tr}h)^2\right]
 +\frac{e^{-4\beta^0}}{2}A^ThA \,,
 \label{eq:curvature-potential}
\end{equation}
where the traces are ordinary matrix traces.  The second term is the magnetic energy of the connection. 

Now, let $r_a$ denote the the $a$th row of $N$ and write $h=\sum_ae^{-2\beta^a}r_ar_a^T$. The magnetic term is then a sum of squares, $A^ThA=\sum_ae^{-2\beta^a}\bigl[(NA)_a\bigr]^2$, which gives the three magnetic walls in Table~\ref{tab:wall-catalogue}. The sphere term is the same as the four-dimensional Bianchi IX potential, and reads
\begin{equation}
 2\operatorname{tr}(h^2)-(\operatorname{tr}h)^2
 =\sum_ae^{-4\beta^a}|r_a|^4
 +\sum_{a\neq b}e^{-2(\beta^a+\beta^b)}\bigl[2(r_a\cdot r_b)^2-|r_a|^2|r_b|^2\bigr].
\end{equation}
The diagonal sum gives the three sphere walls in Table~\ref{tab:wall-catalogue}. The off-diagonal sum gives the cross terms
\begin{equation}
 \begin{aligned}
 &\Bigl[2(n_{12}+n_{13}n_{23})^2
 -(1+n_{12}^2+n_{13}^2)(1+n_{23}^2)\Bigr]
 e^{-2(\beta^0+\beta^1+\beta^2)}\\
 &\quad+\bigl(n_{13}^2-1-n_{12}^2\bigr)e^{-2(\beta^0+\beta^1+\beta^3)}
 +\bigl(n_{23}^2-1\bigr)e^{-2(\beta^0+\beta^2+\beta^3)} .
 \end{aligned}
 \label{eq:curvature-cross-terms}
\end{equation}
Table~\ref{tab:wall-catalogue} and Eq.~\eqref{eq:curvature-cross-terms} together give the full curvature potential~\eqref{eq:curvature-potential}.

\subsection{Building the billiard}
The potential in Eq.~\eqref{eq:full-hamiltonian} consists of three kinds of terms:
\begin{itemize}
\item the twelve wall terms of Table~\ref{tab:wall-catalogue}, whose wall forms all have spacelike normals, $G^{AB}w_Aw_B=2$, and whose coefficients are nonnegative;
\item the three cross terms~\eqref{eq:curvature-cross-terms};
\item the cosmological term $-12L^{-2}e^{-2\sum_A\beta^A}$.
\end{itemize}
We define $\beta^A = \rho_\beta \gamma^A$ where $\rho_\beta^2 = - G_{AB} \beta^A \beta^B$, and hence $\gamma$ comprise directions on a hyperboloid. Each of the wall terms then factorize as $w(\beta) = \rho_\beta w(\gamma)$. As $\rho_\beta \to \infty$, a wall term with a coefficient that is nonzero becomes a hard reflecting wall at $w(\gamma) = 0$.

The cross terms and the cosmological term do not produce walls. Their exponents, which are $\beta^0+\beta^a+\beta^b$ and $\sum_A\beta^A$, respectively, are positive everywhere inside the future light cone of the DeWitt metric, where the billiard motion takes place. These terms therefore decay as $\rho_\beta\to\infty$ in every direction, whatever the sign of their coefficients~\cite{Damour:2002et}. The exception is when an exact symmetry makes a cross term exponent equal to a wall exponent. The two terms must then be combined before taking the limit, because the cross term can cancel the wall---see below. 

\subsection{The horizon data and excluded walls}

When the horizon metric $H$ has a repeated eigenvalue, its eigenbasis is not unique, and the off-diagonal components of $h_{ab}$ that generate the angular walls can depend on which eigenbasis is used. To see this, note that the connection-dependent terms in Eqs.~\eqref{eq:horizon-h-coefficients}--\eqref{eq:horizon-a-coefficients} are the components, in an eigenbasis of $H$, of
\begin{equation}
 A^{(2)}=\frac{H^2A^{\rm H}}{8\det H},
 \qquad
 h^{(4)}\Big|_{A}=\frac{\kappa^2}{16\det H}
 \Bigl[\bigl(A^{\rm H}\cdot HA^{\rm H}\bigr)H-\bigl(HA^{\rm H}\bigr)\bigl(HA^{\rm H}\bigr)^T\Bigr],
 \label{eq:adapted-horizon}
\end{equation}
where $h^{(4)}|_A$ is the connection-dependent part of $h^{(4)}$; the rest of $h^{(4)}$ is diagonal in any eigenbasis. The first term in $h^{(4)}|_A$ is proportional to $H$, so it is also diagonal. The off-diagonal components come entirely from $(HA^{\rm H})(HA^{\rm H})^T$, giving $h^{(4)}_{ab}=-\kappa^2H_aH_bA^{\rm H}_aA^{\rm H}_b/(16\det H)$ for $a\neq b$, as in Eq.~\eqref{eq:horizon-off-diagonal}. Suppose, for example, that $H_1=H_2$ and $A^{\rm H}=(A_1,A_2,0)$ with $A_1A_2\neq0$, so that $h^{(4)}_{12}\neq0$. Any rotation in the $1$--$2$ plane gives another eigenbasis of $H$, and in the one where $A^{\rm H}$ points along the first axis, $A_2=0$ and $h^{(4)}_{12}=0$. The mixing was therefore an artifact of the basis. If instead $H_1\neq H_2$, no such rotation is available, and the mixing is physical.

We therefore use an \emph{adapted} eigenbasis. Within each eigenspace of $H$, the basis is rotated so that the projection of $A^{\rm H}$ onto that eigenspace lies along a single basis vector. In this basis, distinct nonzero components of $A^{\rm H}$ lie in distinct eigenspaces. Their number $k$ is the number of eigenspaces onto which $A^{\rm H}$ has a nonzero projection. Labelling these components first, $A^{\rm H}=(A_1,\dots,A_k,0,\dots,0)$. The invariant
\begin{equation}
 \bigl|A^{\rm H}\times(HA^{\rm H})\bigr|^2
 =\sum_{a<b}(H_a-H_b)^2(A^{\rm H}_a)^2(A^{\rm H}_b)^2
 \label{eq:alignment}
\end{equation}
is positive if and only if $k\ge2$, that is, if and only if $A^{\rm H}$ is not an eigenvector of $H$---this is the criterion~\eqref{eq:criterion}. 

\subsubsection{(i) Discrete symmetries remove walls.}

We now have enough technology to explain how discrete symmetries of the horizon data, and the solution, remove walls. We will illustrate this explicitly for the following transformation,
\begin{equation}
 \mathcal P_3:\quad(\dn T,\sigma^1,\sigma^2,\sigma^3)\mapsto(-\dn T,-\sigma^1,-\sigma^2,\sigma^3).
\end{equation}
which preserves the structure constants. Due to this, along with the fact that the time reversal is also just a coordinate change, this transformation maps any solution of the form~\eqref{eq:ansatz} to another solution of the same form. To find the new metric functions, apply the transformation to Eq.~\eqref{eq:ansatz}. Under this transformation, in $h_{ab}\sigma^a\sigma^b$, only the terms containing exactly one factor of $\sigma^3$ change sign. In the cross-term $(\dn T + A_a \sigma^a)^2$, an overall sign can be pulled out leaving a sign change only in the $A_3 \sigma^3$ term. Hence the transformation acts as
\begin{equation}
 A_3\mapsto-A_3,\qquad h_{13}\mapsto-h_{13},\qquad h_{23}\mapsto-h_{23},
\end{equation}
and leaves $F$, $A_1$, $A_2$, $h_{12}$ and the diagonal components of $h$ unchanged.

Now consider the horizon data. Since $F$ and the diagonal $H$ are unchanged, $\mathcal P_3$ maps $(\kappa,H,A^{\rm H}_1,A^{\rm H}_2,A^{\rm H}_3)$ to $(\kappa,H,A^{\rm H}_1,A^{\rm H}_2,-A^{\rm H}_3)$. If $A^{\rm H}_3=0$, the transformed solution therefore has the same horizon data as the original, and by uniqueness the two coincide. Each function that $\mathcal P_3$ reverses is then equal to minus itself, so it vanishes: If $A^{\rm H}_3=0$ in the adapted eigenbasis, then $A_3$, $h_{13}$ and $h_{23}$ vanish identically, both outside the horizon and inside it. Writing $\mathcal P_c$ for the analogous transformation about axis $c$, the same holds for any $c$ with $A^{\rm H}_c=0$.

As an explicit example, take $k=2$, so that $A^{\rm H}=(A_1,A_2,0)$ with $H_1\neq H_2$. The statement above gives $A_3=h_{13}=h_{23}=0$, and Eq.~\eqref{eq:angular-mixing} then gives $n_{13}=n_{23}=0$, while $n_{12}=h_{12}/h_{11}$ is not constrained. Substituting these results into Table~\ref{tab:wall-catalogue}, one sees that four wall coefficients vanish identically:
\begin{equation}
 \begin{gathered}
 C_{c_{012}}=\tfrac12A_3^2=0,\qquad
 \Omega_{03}=\dot A_3-n_{23}\dot A_2+(n_{12}n_{23}-n_{13})\dot A_1=0,\\
 \Omega_{13}=\dot n_{13}-n_{23}\dot n_{12}=0,\qquad
 \Omega_{23}=\dot n_{23}=0 .
 \end{gathered}
\end{equation}
The surviving coefficients reduce to
\begin{equation}
 \begin{gathered}
 C_{c_{023}}=\tfrac12(A_1+n_{12}A_2)^2,\qquad
 C_{c_{013}}=\tfrac12A_2^2,\qquad
 C_{c_{123}}=\tfrac12(1+n_{12}^2)^2,\qquad
 C_{c_{213}}=C_{c_{312}}=\tfrac12,\\
 \Omega_{01}=\dot A_1,\qquad
 \Omega_{02}=\dot A_2-n_{12}\dot A_1,\qquad
 \Omega_{12}=\dot n_{12},
 \end{gathered}
\end{equation}
and the cross terms~\eqref{eq:curvature-cross-terms} reduce to
$(n_{12}^2-1)e^{-2(\beta^0+\beta^1+\beta^2)}-(1+n_{12}^2)e^{-2(\beta^0+\beta^1+\beta^3)}-e^{-2(\beta^0+\beta^2+\beta^3)}$.

For $k=1$, $A^{\rm H}=(A_1,0,0)$, the statement applies with $c=2$ and with $c=3$. Together these give $A_2=A_3=0$ and $h_{12}=h_{13}=h_{23}=0$, so $h$ is diagonal and $N=I$. The only surviving magnetic and symmetry wall coefficients are then $C_{c_{023}}=\tfrac12A_1^2$ and $C_{s_{01}}$. All three sphere coefficients equal $\tfrac12$, and all coefficients in the cross term equal $-1$. For $k=0$ the statement applies for every $c$: the connection vanishes identically and $h$ stays diagonal, so only the sphere walls and the cross terms remain. These results are collected in Table~\ref{tab:zero-walls}.

\begin{table}[t]
\centering\small
\renewcommand{\arraystretch}{1.2}
\begin{tabular}{@{}clll@{}}\toprule
$k$ & $A^{\rm H}$ & Fields vanishing identically & Coefficients vanishing identically\\\midrule
3 & $(A_1,A_2,A_3)$ & none & none\\
2 & $(A_1,A_2,0)$ & $A_3$, $h_{13}$, $h_{23}$ \ ($n_{13}=n_{23}=0$) & $c_{012}$; \ $s_{03},s_{13},s_{23}$\\
1 & $(A_1,0,0)$ & $A_2$, $A_3$, $h_{12}$, $h_{13}$, $h_{23}$ \ ($N=I$) & $c_{013},c_{012}$; \ $s_{02},s_{03},s_{12},s_{13},s_{23}$\\
0 & $0$ & $A_a$, $h_{12}$, $h_{13}$, $h_{23}$ & $c_{023},c_{013},c_{012}$; \ all six $s_{AB}$\\\bottomrule
\end{tabular}
\caption{Coefficients removed by the discrete symmetries $\mathcal P_c$, in the adapted eigenbasis with nonzero components of $A^{\rm H}$ listed first. By Eq.~\eqref{eq:horizon-coefficients}, every coefficient not listed is nonzero just inside the horizon, so the list is complete. Coefficients of sphere curvature walls never vanish.}
\label{tab:zero-walls}
\end{table}

The converse follows from the interior continuation of the near-horizon series. Working with $\kappa=1$, we have then $q_{00}=t^2+O(t^4)$, and Eqs.~\eqref{eq:horizon-off-diagonal}, \eqref{eq:horizon-a-coefficients} and~\eqref{eq:angular-mixing} give
\begin{equation}
 \dot A_a=-\frac{H_a^2A^{\rm H}_a}{4\det H}\,t+O(t^3),
 \qquad
 n_{ab}=-\frac{H_bA^{\rm H}_aA^{\rm H}_b}{16\det H}\,t^4+O(t^6)\quad \text{for } a<b.
\end{equation}
As an example, take $s_{12}$. Using $V^2=t^2\det H+O(t^4)$ and $e^{4(\beta^2-\beta^1)}=H_1^2/H_2^2+O(t^2)$ in Eq.~\eqref{eq:symmetry-coefficients} gives
\begin{equation}
 C_{s_{12}}=\frac{t^2\det H}{2}\,\frac{H_1^2}{H_2^2}
 \left(\frac{H_2A^{\rm H}_1A^{\rm H}_2}{4\det H}\,t^3\right)^2+O(t^{10})
 =\frac{H_1^2(A^{\rm H}_1A^{\rm H}_2)^2}{32\det H}\,t^8+O(t^{10}).
\end{equation}
In the same way, with no sums over displayed indices,
\begin{equation}
 \begin{aligned}
 C_{c_{0bc}}&=\frac12(A^{\rm H}_a)^2+O(t^2)\,&
 C_{c_{abc}}&=\frac12+O(t^8)\,,\\
 C_{s_{0a}}&=\frac{H_a^2(A^{\rm H}_a)^2}{32\det H}\,t^8+O(t^{10})\,&
 C_{s_{ab}}&=\frac{H_a^2(A^{\rm H}_aA^{\rm H}_b)^2}{32\det H}\,t^8+O(t^{10})\quad \text{for } a<b \,.
 \end{aligned}
 \label{eq:horizon-coefficients}
\end{equation}
Thus each nonzero adapted component $A^{\rm H}_a$ switches on the magnetic wall $c_{0bc}$ and the electric wall $s_{0a}$, and each pair of nonzero components switches on the angular wall $s_{ab}$. In the adapted basis these pairs lie in distinct eigenspaces, so the angular mixing is physical rather than a choice of basis. Every coefficient not listed in Table~\ref{tab:zero-walls} therefore has a nonzero leading term, and that table is complete.

\subsubsection{(ii) Continuous symmetries cancel walls.}
The cross terms are normally subleading because their exponents differ from those of the walls. If a symmetry forces two scale factors to be equal, however, a cross term exponent can coincide with wall exponents. The terms must then be combined, and they can cancel. This happens only when $k\le1$, so by (i) the angular metric is diagonal, $N=I$, and all three cross term coefficients equal $-1$. The sphere part of Eq.~\eqref{eq:curvature-potential} is then
\begin{equation}
 \frac{e^{-2\beta^0}}{2}
 \left[\sum_{a}e^{-4\beta^a}-2\sum_{a<b}e^{-2(\beta^a+\beta^b)}\right].
 \label{eq:diagonal-sphere-potential}
\end{equation}
We now consider each possibility in turn. 

\emph{Biaxial horizon.} Suppose that $H_2=H_3$ and that $A^{\rm H}=(A_1,0,0)$ points along the 1 axis, and we allow for $A_1=0$. Consider a rotation of the frame by any angle about the 1 axis. This is a change of coordinates, so it maps solutions to solutions. It leaves $H$ unchanged, because $H$ is proportional to the identity in the 2--3 plane, and it leaves $A^{\rm H}$ unchanged, because $A^{\rm H}$ lies along the rotation axis. By uniqueness, every such rotation is therefore a symmetry of the solution. The 2--3 block of $h$ must then be invariant under all rotations in that plane, which forces $h_{22}=h_{33}$, and hence $\beta^2=\beta^3$ for all $t$. This then forces coincidence of some exponents: two sphere walls and one cross term now share the exponent $\beta^0+2\beta^2$, and they cancel exactly:
\begin{equation}
 \Bigl[\frac12e^{-2(\beta^0+2\beta^2)}+\tfrac12e^{-2(\beta^0+2\beta^3)}-e^{-2(\beta^0+\beta^2+\beta^3)}\Bigr]_{\beta^2=\beta^3}=0.
\end{equation}
The remaining sphere potential is
\begin{equation}
 \frac12e^{-2(\beta^0+2\beta^1)}-2e^{-2(\beta^0+\beta^1+\beta^2)},
 \label{eq:biaxial-potential}
\end{equation}
so the only positive sphere wall is $c_{123}$. The walls $c_{213}$ and $c_{312}$ are absent.

\emph{Round horizon.} If $H$ is proportional to the identity and $A^{\rm H}=0$, every rotation leaves the horizon data unchanged. The same argument then gives $h\propto I$ for all $t$, so $\beta^1=\beta^2=\beta^3$. All six sphere terms now share one exponent, and the sphere potential becomes
\begin{equation}
 \Bigl(\frac32-3\Bigr)e^{-2(\beta^0+2\beta^1)}=-\frac32e^{-2(\beta^0+2\beta^1)}.
 \label{eq:round-potential}
\end{equation}
This is negative, so no positive curvature wall remains.

The cancellation relies on the rotation symmetry, and hence on all of the horizon data being symmetric, not just $H$. Suppose $H_2=H_3$, but $A^{\rm H}$ has a nonzero component in the 2--3 plane as well as along the 1 axis. That component picks out a direction in the plane, so rotations about the 1 axis no longer leave the horizon data unchanged. Nothing then forces $\beta^2=\beta^3$, and no cancellation occurs. Rotating within the 2--3 plane brings these data to the form $A^{\rm H}=(A_1,A_2,0)$ with $H_1\neq H_2$, which is the $k=2$ class.

\subsubsection{(iii) Redundant walls}

In the sharp wall limit, each wall that is present confines the motion to $w(\beta)\ge0$, so the billiard chamber is the region where all of these inequalities hold simultaneously. Some of these inequalities are implied by others. Suppose a wall form can be written as $w=\sum_i\lambda_iw_i$, with every $\lambda_i\ge0$ and every $w_i$ active. Inside the chamber each $w_i\ge0$, so $w\ge0$ automatically: the trajectory reaches one of the $w_i$, and reflects there, before it could reach $w$. Such a wall never bounds the chamber, and we call it redundant. 

To apply this, note that every wall form is a nonnegative combination of the four forms $c_{023}$, $s_{01}$, $s_{12}$ and $s_{23}$:
\begin{equation}
 \begin{aligned}
 c_{123}&=c_{023}+s_{01}, & c_{013}&=c_{023}+s_{12}, & s_{02}&=s_{01}+s_{12},\\
 c_{213}&=c_{023}+s_{01}+2s_{12}, & c_{012}&=c_{023}+s_{12}+s_{23}, & s_{03}&=s_{01}+s_{12}+s_{23},\\
 c_{312}&=c_{023}+s_{01}+2s_{12}+2s_{23}, & & & s_{13}&=s_{12}+s_{23}.
 \end{aligned}
 \label{eq:wall-reductions}
\end{equation}
A wall is therefore redundant whenever every form on the right-hand side of its expansion is present. Conversely, a wall that is the only one present containing a given $\beta^a$ with positive coefficient cannot be a combination of the others, and so it is never redundant. Applying these two observations to the walls that survive (i) and (ii) gives the following.
\begin{itemize}
\item $k=0$. Only the three sphere walls are present, and their expansions all require $c_{023}$ and $s_{01}$, which are absent. With distinct eigenvalues, each sphere wall is the only one containing its own $\beta^a$, so all three are dominant: $c_{123},c_{213},c_{312}$. With $H_1\neq H_2=H_3$, mechanism (ii) leaves only $c_{123}$. For a round horizon, no wall with positive coefficient remains.
\item $k=1$. The nonvanishing connection component adds $c_{023}$ and $s_{01}$. This makes $c_{123}=c_{023}+s_{01}$ redundant. The expansions of $c_{213}$ and $c_{312}$ require $s_{12}$, which is absent, and these two walls are the only ones containing $\beta^2$ and $\beta^3$ respectively, so they remain. The dominant walls are $c_{023},s_{01},c_{213},c_{312}$. If $H_2=H_3$, mechanism (ii) removes $c_{213}$ and $c_{312}$, leaving $c_{023},s_{01}$.
\item $k=2$. A second component of the connection adds $c_{013}$, $s_{02}$ and $s_{12}$. Now $c_{013}$, $s_{02}$, $c_{123}$ and $c_{213}$ all have expansions using only $c_{023}$, $s_{01}$ and $s_{12}$, so they are redundant. The expansion of $c_{312}$ requires $s_{23}$, which is absent, and $c_{312}$ is the only wall containing $\beta^3$, so it remains. The dominant walls are $c_{023},s_{01},s_{12},c_{312}$.
\item $k=3$. A third component adds $c_{012}$, $s_{03}$, $s_{13}$ and $s_{23}$. All four basic forms are now present, so every other wall is redundant. The dominant walls are $c_{023},s_{01},s_{12},s_{23}$, as in Eq.~\eqref{eq:ae4}.
\end{itemize}
These results are summarized in Table~\ref{tab:billiards}.

\section{Numerical methods and validation}
\label{app:numerics}

Our numerical solutions were constructed in two steps. For a given set of horizon data, we use the near-horizon expansion to initialize the exterior integration, where we integrate toward the conformal boundary, where we check that the exterior is physically admissible and read off the boundary metric. The same horizon data is then used to initialize an interior integration, which we follow toward the singularity. At every stage we solve the full Einstein equations, $R_{\mu\nu}=-4L^{-2}g_{\mu\nu}$, for the ten components of the metric $q_{AB}$ or an equivalent set of variables. All integrations use the adaptive \texttt{Extrapolation} method of \texttt{NDSolve} in Mathematica, in arbitrary-precision arithmetic, with $L=\kappa=1$. 
 
To check the consistency of these solutions, we used two independent numerical codes: one is written by us and the other one is written by OpenAI's GPT Astra under our direction. We found that both codes give consistent numerical solutions with the imposed horizon data in \eqref{eq:num-example}.

\subsection{Constructing solutions}

\paragraph{Shooting from the horizon.}
Because $q_{AB}$ is degenerate at the horizon, the equations cannot be integrated from $\rho=0$. We therefore extend the near-horizon series for $q$ to order $\rho^{2m}$ by substituting it into the evolution equations and solving for the coefficients order by order; each order is a linear system, and no new constants appear. The series supplies $q$ and $\partial_\rho q$ at a small $\rho_0>0$, where the exterior integration begins. The interior is initialized from the same coefficients with $\rho^2\to-t^2$, at a small $t_0>0$. A smaller $\rho_0$ or $t_0$ reduces the truncation error of the series but amplifies rounding errors in the nearly degenerate metric, so we vary the series order, starting point, and precision together and compare the solutions away from the horizon.

\paragraph{Exterior.}
Near the boundary $q$ grows as $e^{2\rho/L}$. To remove this growth we evolve $\mathsf Q=e^{-2\rho/L}q$ and $\mathsf P=\partial_\rho\mathsf Q$, which approach finite limits. The metric evolution equation becomes
\begin{equation}
\begin{aligned}
 \partial_\rho\mathsf P={}&2e^{-2\rho/L}\operatorname{Ric}(\mathsf Q)
 +\mathsf P\mathsf Q^{-1}\mathsf P
 -\frac12\operatorname{tr}(\mathsf Q^{-1}\mathsf P)\,\mathsf P-\frac1L\operatorname{tr}(\mathsf Q^{-1}\mathsf P)\,\mathsf Q
 -\frac4L\mathsf P .
\end{aligned}
 \label{eq:num-exterior}
\end{equation}
where $\operatorname{Ric}(\mathsf Q)$ is the  Ricci tensor of ${\mathsf Q}$ with lowered indices, which is unchanged by constant rescalings of the metric. We integrate to $\rho=16L$. An integrated solution is an admissible exterior only if $F>0$, $h_{ab}>0$ and $v^2=F^2h^{ab}A_aA_b<1$ for all $\rho>0$.  Given $h>0$, the last condition is equivalent to positivity of $h-F^2AA^T$. We check these conditions at the integration nodes and other sampled points, and locate any minima between samples from sign changes of the derivatives, so that a dip between samples is not missed. The near-horizon series covers the region below the starting point, and the boundary fit covers the region beyond the end of the integration, so together these checks give numerical evidence that the whole exterior is admissible. The boundary metric is obtained by fitting each component of $\mathsf Q$ to a constant plus terms in $e^{-2\rho/L}$, $e^{-4\rho/L}$ and $(\rho/L)e^{-4\rho/L}$, the last being the logarithmic term expected for a curved boundary, and comparing the windows $13\le\rho/L\le15$ and $14\le\rho/L\le16$. The settings and diagnostics for the solution with horizon data
\begin{equation}
 H=2(399/400)^{1/3}\operatorname{diag}(9/5,1,5/9),\qquad
 A^{\rm H}=(24/25,12/25,18/25),
 \label{eq:num-example}
\end{equation}
which is used in the Letter, are listed in Tables~\ref{tab:num-runs} and~\ref{tab:num-diagnostics}.

\paragraph{Interior variables.} The scale factors separate exponentially as the singularity is approached, and evolving $q$ directly could therefore lose exponentially small terms that may become important later. In the interior we therefore evolve the Iwasawa variables, written as $q=E^TE$ with $E=DU$, $D=\operatorname{diag}(e^{a_A})$, $U$ upper triangular with unit diagonal, and $a_A=-\beta^A$, together with the extrinsic curvature in the orthonormal frame defined by $E$,
\begin{equation}
 \widehat K=\frac12(E^{-1})^{T}\,(\partial_tq)\,E^{-1},\qquad
 \Theta=-V\operatorname{tr}\widehat K,\qquad V=\sqrt{\det q} \, .
\end{equation}
The interior first expands and then contracts. Through the volume maximum we evolve $a_A$, $U$ and $V\widehat K$ in $\eta$ time, $\dn t=V\dn\eta$. At $\widehat K = 0$, we record the maximum volume $V_{\rm max}$ for the solution. Once $\Theta$ reaches $0.1$ after the maximum, we switch to the volume clock and normalized expansion,
\begin{equation}
 x=\ln(V_{\rm max}/V),\qquad B=-\widehat K/\operatorname{tr}\widehat K,\qquad \mu=\ln\Theta,
\end{equation}
where $V_{\rm max}$ is the volume at the maximum. Thus $x$ differs from $-\ln V$ by a constant, $\operatorname{tr}B=-1$, and the Iwasawa exponents are $p_A=-B_{AA}=\dn\beta^A/\dn x$. The numerical values of $x$ below use this convention. With $\mathfrak R=\widehat{\operatorname{Ric}}/(\operatorname{tr}\widehat K)^2$ the normalized spatial Ricci tensor, $\lambda_\Lambda=1/(L\operatorname{tr}\widehat K)^2$, and $\mathcal O(X)$ the upper-triangular matrix with diagonal $X_{AA}$ and entries $2X_{AB}$ for $A<B$, the Einstein equations become
\begin{equation}
\begin{aligned}
 \partial_xa_A&=B_{AA},&
 \partial_xU&=D^{-1}\bigl[\mathcal O(B)-\operatorname{diag}B\bigr]DU,\\
 \partial_x\mu&=\operatorname{tr}\mathfrak R+16\lambda_\Lambda,&
 \partial_xB&=[\mathcal O(B),B]-\mathfrak R-4\lambda_\Lambda I-B\,\partial_x\mu \, .
\end{aligned}
\label{eq:num-interior}
\end{equation}
We use two further techniques to keep exponentially small quantities accurate. First, between bounces the off-diagonal $B_{AB}$ become extremely small, so we evolve $\Pi_{AB}=\Theta e^{a_A-a_B}B_{AB}$ with $A<B$ in their place. This removes the exponential suppression of the off-diagonal variables. This combination is further natural because the $\Pi_{AB}$ are directly related to the coefficients of the symmetry walls, $C_s \propto \Pi^2$, and we therefore expect (but do not assume) them to approach constants.  Second, $\mathfrak R$ is computed without forming $q$ or its inverse. On a homogeneous space, the Ricci tensor in an orthonormal frame is quadratic in the structure constants of that frame. We therefore introduce the structure constants of the orthonormal frame $E=DU$, rescaled by $1/|\operatorname{tr}\widehat K|=e^{\sum_Aa_A-\mu}$,
\begin{equation}
 \mathcal W^A{}_{BC}=e^{\sum_Da_D+a_A-a_B-a_C-\mu}\,U^A{}_D(U^{-1})^E{}_B(U^{-1})^F{}_C\,C^D{}_{EF},
\end{equation}
and obtain $\mathfrak R$ from Eq.~\eqref{eq:orbit-ricci} with $q$ replaced by the identity and $C$ by $\mathcal W$, where $C^D{}_{EF}$ are the structure constants of Appendix~A. Each component of $\mathcal W$ is a single exponential times a polynomial in $U$, so no cancellations between exponentially different terms occur. The evolved state therefore has 22 components, $(a_A,U_{AB},B_{AA},\Pi_{AB},\mu,t)$. These 22 variables encode $q$ and $\partial_tq$ (20 numbers) together with $t$, subject to the identities $\operatorname{tr}B=-1$ and $\sum_Aa_A=\ln V_*-x$, which are not imposed but monitored.  The equations for $B_{AA}$ and $\Pi_{AB}$ follow from Eq.~\eqref{eq:num-interior}, and $\partial_xt=e^{\sum_Aa_A-\mu}$. 

\subsection{Integration settings and validation}

Our standard interior integrations start independently from the horizon and reach $x=10^8$, using 80 and 100 digits of working precision. Each computes its own horizon series, with the order and offset listed in Table~\ref{tab:num-runs}. Its precision and error goals are retained throughout. The adaptive step cap is large during Kasner epochs and reduced near interactions, guided by the wall amplitudes and their local rates of change. This controls numerical resolution without modifying the equations. 

\begin{table}[htbp]
\centering\small
\setlength{\tabcolsep}{6pt}
\renewcommand{\arraystretch}{1.15}
\begin{tabular}{@{}lrrlll@{}}\toprule
Integration & Digits & Goals & Series & Starting offset & Step cap\\\midrule
Exterior & 100 & 46 & $\rho^{24}$ & $\rho_0=10^{-4}$ & $\Delta\rho\le1/12$\\
Interior & 80 & 35 & $t^{24}$ & $t_0=10^{-5}$ & $\Delta x\le30\,000$ / $4$\\
Interior check & 100 & 45 & $t^{20}$ & $t_0=10^{-6}$ & $\Delta x\le7\,500$ / $2$\\\bottomrule
\end{tabular}
\caption{Standard numerical settings for Eq.~\eqref{eq:num-example}. ``Goals'' denotes both the accuracy and precision goals, and ``Series'' the order retained in the near-horizon $q$. The exterior reaches $\rho=16$ and both interiors reach $x=10^8$. The two interior step caps apply during Kasner epochs and near interactions, respectively. The smaller cap is also used for $x\le25$.}
\label{tab:num-runs}
\end{table}

The Hamiltonian and momentum constraints are monitored during the integration.
In the interior, the Hamiltonian residual is divided by the positive scale
$(\operatorname{tr}B)^2+\operatorname{tr}(B^2)+\|\mathfrak R\|_F
+12\lambda_\Lambda$, where
$\|\mathfrak R\|_F=(\sum_{A,B}\mathfrak R_{AB}^{\,2})^{1/2}$ is the Frobenius norm in the spatial orthonormal frame, and each momentum residual by the sum of the absolute values of its terms. These normalizations measure constraint violations relative to the sizes of the terms that should cancel. As an independent check, we reconstruct the full five-dimensional Ricci tensor from derivatives of the stored interpolants, rather than substituting the evolution equations, and normalize the interior Einstein residual by $(\operatorname{tr}\widehat K)^2$, which provides a local curvature scale.

\begin{figure}[htbp]
\centering
\includegraphics[width=0.88\textwidth]{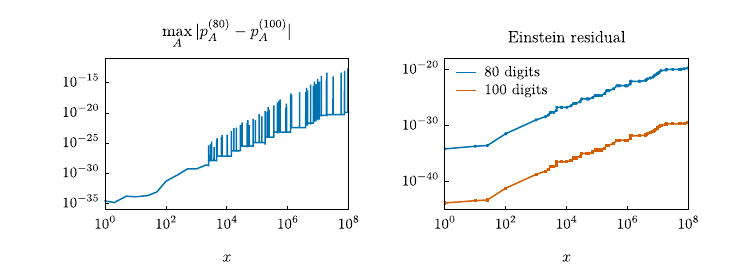}
\caption{Numerical validation of the two independent integrations. Left: the largest componentwise difference in $p_A$ at common $x$. Right: the largest absolute orthonormal component of $R_{\mu\nu}+4L^{-2}g_{\mu\nu}$, divided by $(\operatorname{tr}\widehat K)^2$, reconstructed from interpolant derivatives at the sampled points. Both axes are logarithmic and the plots show $1\le x\le10^8$. Lines join the samples to guide the eye.}
\label{fig:num-evolution-validation}
\end{figure}

In  Table~\ref{tab:num-diagnostics} we report the largest sampled residuals and differences. Comparing the two interior solutions at common $x$ gives agreement in $p_A$ to $2.5\times10^{-13}$ over $0\le x\le10^8$. Both resolve the same ordered sequence of 52 isolated transitions for $x\ge2000$, with their locations agreeing to $4.7\times10^{-12}$ in $x$. Figure~\ref{fig:num-evolution-validation} shows the differences in Iwasawa exponents and independently evaluated residuals of the Einstein equation.

\begin{table}[htbp]
\centering\small
\setlength{\tabcolsep}{6pt}
\renewcommand{\arraystretch}{1.15}
\begin{tabular}{@{}lrr@{}}\toprule
Diagnostic & \multicolumn{2}{c}{Largest sampled value}\\\midrule
\multicolumn{3}{@{}l}{\emph{Exterior}}\\
\quad Hamiltonian residual ($\times L^2$) & \multicolumn{2}{r}{$7.6\times10^{-46}$}\\
\quad Momentum residual ($\times L$) & \multicolumn{2}{r}{$1.5\times10^{-50}$}\\
\quad Independent Einstein residual ($\times L^2$) & \multicolumn{2}{r}{$5.6\times10^{-33}$}\\
\quad Change in boundary metric between fitting windows & \multicolumn{2}{r}{$3.4\times10^{-35}$}\\
\quad $v^2$ & \multicolumn{2}{r}{$0.229348$}\\[2pt]
\emph{Interior, $0\le x\le10^8$} & 80 digits & 100 digits\\
\quad Hamiltonian residual & $1.1\times10^{-20}$ & $2.0\times10^{-30}$\\
\quad Relative momentum residual & $8.0\times10^{-33}$ & $1.1\times10^{-43}$\\
\quad Independent Einstein residual & $2.2\times10^{-20}$ & $3.9\times10^{-30}$\\[2pt]
\multicolumn{3}{@{}l}{\emph{Comparison of the two interior integrations}}\\
\quad Difference in $p_A$ & \multicolumn{2}{r}{$2.4\times10^{-13}$}\\
\quad Difference in $\beta^A$ & \multicolumn{2}{r}{$9.6\times10^{-13}$}\\
\quad Difference in transition locations $x$ & \multicolumn{2}{r}{$4.7\times10^{-12}$}\\\bottomrule
\end{tabular}
\caption{Diagnostics for Eq.~\eqref{eq:num-example}. The exterior constraint and causality checks use 2304 points and the independent check of the Einstein equations uses five points. Interior constraints are sampled at all stored integration nodes, interval midpoints and located transitions, amounting to 119991 points for the 80-digit run and 226445 for the 100-digit run. The independent check of the Einstein equation uses 65 points per run, including the 52 isolated transitions at $x\ge2000$.}
\label{tab:num-diagnostics}
\end{table}

As a further check, we compare with slower integrations at 160 and 130 digits, with accuracy and precision goals of 70 and 55 and step caps $\Delta x\le5$ and $3$, respectively. These start from the horizon with the same respective series orders and offsets as in Table~\ref{tab:num-runs} and reach $x=2\,523\,750$. At 804 comparison points, including points near every isolated deep transition in this interval, the 80- and 100-digit solutions agree with the corresponding slower integrations in $p_A$ to $3.1\times10^{-17}$ and $5.6\times10^{-27}$. These slower integrations provide consistency checks of the larger steps taken during the Kasner epochs. This comes at a cost, though: the slower integrations through $x=2\,523\,750$ occupy approximately 60~GB. The two standard interpolants through $x=10^8$ together occupy 1.4~GB.

\subsection{Jacobian at fixed horizon area and DeWitt radius}

To test the openess of the set of billiard trajectories accessible to regular horizon data, we evaluate the Jacobian~\eqref{eq:openness-jacobian} from the five independent horizon parameters to $(\gamma^1,\gamma^2,\gamma^3,p_1,p_2)$, holding $\det H$ and $\rho_\beta$ fixed. Every perturbed solution is initialized independently at the horizon and evolved to $\rho_\beta=6779.42497522305\ldots$, which is reached at $x=8500$ by the solution~\eqref{eq:num-example}. For each perturbed solution, we find the value of $x$ at which it reaches the prescribed $\rho_\beta$. This value generally differs between solutions, so comparing them at the same $x$ would not compare them at the same billiard depth. We choose the above value of $\rho_\beta$ because there the solution is in a long, well-resolved Kasner epoch, but not too deep into the interior that constructing multiple solutions becomes too time consuming. During the Kasner epoch, the relations $\sum_A p_A=\sum_A p_A^2=1$ leave only two independent exponents. Since $p_0\neq p_3$ in this Kasner epoch, small changes in $p_1$ and $p_2$ uniquely determine the resulting small changes in $p_0$ and $p_3$.

We compute the Jacobian by numerical differencing, using the logarithmic coordinates $(\ln H_1,\ln H_2,\ln H_3,\ln A_1^{\rm H},\ln A_2^{\rm H}, \ln A_3^{\rm H})$. Starting from the horizon data in Eq.~\eqref{eq:num-example}, a displacement $\epsilon v$ in these coordinates means replacing $H_a\to H_a e^{\epsilon v_a}$ and $A_a^{\rm H}\to A_a^{\rm H}e^{\epsilon v_{a+3}}$, for $a=1,2,3$. For the displacement directions $v$, we choose each of the following five orthonormal vectors:
\begin{equation}
\begin{gathered}
 (1,-1,0,0,0,0)/\sqrt2,\qquad (1,1,-2,0,0,0)/\sqrt6,\\
 (0,0,0,1,0,0),\qquad (0,0,0,0,1,0),\qquad (0,0,0,0,0,1).
\end{gathered}
\label{eq:num-log-basis}
\end{equation}
The first two vary the horizon shape while preserving $\det H$, and the remaining three vary the connection while leaving $H$ unchanged. All five perturbations therefore preserve the horizon area.

For each direction, we integrate the solutions with displacements $\epsilon=\pm \delta$ and $\pm \delta/2$, where $\delta=10^{-8}$, giving twenty perturbed solutions. These use 70 working digits, accuracy and precision goals of 25, horizon series to order $t^{12}$ or $t^{16}$, and $t_0=10^{-7}$. For an interior parameter $f$, let $f(\epsilon)$ denote its value at the prescribed depth for horizon data displaced by $\epsilon v$. We estimate its derivative at the unperturbed solution using central differences. We then use the chain rule to obtain derivatives with respect to the individual horizon parameters $(H_1,H_2,A_1^{\rm H},A_2^{\rm H},A_3^{\rm H})$. Applying this procedure to all five interior parameters gives the columns of the Jacobian defined in the main text. 

The central difference formula includes an error term quadratic in $\delta$. If we half the step size, this reduces the error by a known factor of four. Given those two evaluations, Richardson extrapolation can be applied to eliminate the quadratic error entirely: 
\begin{equation}
 J_{\rm ext}=\frac{4J_{\delta/2}-J_\delta}{3}=J+O(\delta^4) \,,
\end{equation}
provided integration and rounding errors are sufficiently small. We then obtain
\begin{equation}
\begin{gathered}
 \det J_{\rm ext}=1.58054\times10^{-3},\\
 \sigma_{\min}(J_{\delta/2})=1.04585\times10^{-4},\qquad
 \sigma_{\min}(J_{\rm ext})=1.04582\times10^{-4},
\end{gathered}
\label{eq:num-jacobian-check}
\end{equation}
where $\sigma_{\min}$ denotes the smallest singular value, which measures the weakest response of the interior data to a unit variation of the horizon parameters. The fact that the smallest singular value is nonzero, and further that it is stable under extrapolation,  supports the conclusion that variations in the horizon data resolve all five directions in the billiard.  To check that numerical integration errors are smaller than the observed step-size dependence, we repeat the calculation at $h/2$ with higher precision and tighter tolerances. This changes the smallest singular value by only $1.27\times10^{-10}$ fractionally, compared with the $2.8\times10^{-5}$ change under Richardson extrapolation. The precision dependence is therefore much smaller than the finite-difference correction. 

\vspace{50pt}

\twocolumngrid
\newpage
\clearpage
\bibliography{bkl-refs}

\begin{thebibliography}{46}%
\makeatletter
\providecommand \@ifxundefined [1]{%
 \@ifx{#1\undefined}
}%
\providecommand \@ifnum [1]{%
 \ifnum #1\expandafter \@firstoftwo
 \else \expandafter \@secondoftwo
 \fi
}%
\providecommand \@ifx [1]{%
 \ifx #1\expandafter \@firstoftwo
 \else \expandafter \@secondoftwo
 \fi
}%
\providecommand \natexlab [1]{#1}%
\providecommand \enquote  [1]{``#1''}%
\providecommand \bibnamefont  [1]{#1}%
\providecommand \bibfnamefont [1]{#1}%
\providecommand \citenamefont [1]{#1}%
\providecommand \href@noop [0]{\@secondoftwo}%
\providecommand \href [0]{\begingroup \@sanitize@url \@href}%
\providecommand \@href[1]{\@@startlink{#1}\@@href}%
\providecommand \@@href[1]{\endgroup#1\@@endlink}%
\providecommand \@sanitize@url [0]{\catcode `\\12\catcode `\$12\catcode
  `\&12\catcode `\#12\catcode `\^12\catcode `\_12\catcode `\%12\relax}%
\providecommand \@@startlink[1]{}%
\providecommand \@@endlink[0]{}%
\providecommand \url  [0]{\begingroup\@sanitize@url \@url }%
\providecommand \@url [1]{\endgroup\@href {#1}{\urlprefix }}%
\providecommand \urlprefix  [0]{URL }%
\providecommand \Eprint [0]{\href }%
\providecommand \doibase [0]{https://doi.org/}%
\providecommand \selectlanguage [0]{\@gobble}%
\providecommand \bibinfo  [0]{\@secondoftwo}%
\providecommand \bibfield  [0]{\@secondoftwo}%
\providecommand \translation [1]{[#1]}%
\providecommand \BibitemOpen [0]{}%
\providecommand \bibitemStop [0]{}%
\providecommand \bibitemNoStop [0]{.\EOS\space}%
\providecommand \EOS [0]{\spacefactor3000\relax}%
\providecommand \BibitemShut  [1]{\csname bibitem#1\endcsname}%
\let\auto@bib@innerbib\@empty
\bibitem [{\citenamefont {Belinsky}\ \emph {et~al.}(1970)\citenamefont
  {Belinsky}, \citenamefont {Khalatnikov},\ and\ \citenamefont
  {Lifshitz}}]{Belinsky:1970ew}%
  \BibitemOpen
  \bibfield  {author} {\bibinfo {author} {\bibfnamefont {V.~A.}\ \bibnamefont
  {Belinsky}}, \bibinfo {author} {\bibfnamefont {I.~M.}\ \bibnamefont
  {Khalatnikov}},\ and\ \bibinfo {author} {\bibfnamefont {E.~M.}\ \bibnamefont
  {Lifshitz}},\ }\bibfield  {title} {\bibinfo {title} {{Oscillatory approach to
  a singular point in the relativistic cosmology}},\ }\href
  {https://doi.org/10.1080/00018737000101171} {\bibfield  {journal} {\bibinfo
  {journal} {Adv. Phys.}\ }\textbf {\bibinfo {volume} {19}},\ \bibinfo {pages}
  {525} (\bibinfo {year} {1970})}\BibitemShut {NoStop}%
\bibitem [{\citenamefont {Chitre}(1972)}]{Chitre:1972}%
  \BibitemOpen
  \bibfield  {author} {\bibinfo {author} {\bibfnamefont {D.~M.}\ \bibnamefont
  {Chitre}},\ }\emph {\bibinfo {title} {Investigation of Vanishing of a Horizon
  for Bianchi Type IX (the Mixmaster) Universe}},\ \href
  {https://doi.org/10.13016/qgai-gtpd} {Ph.D. thesis},\ \bibinfo  {school}
  {University of Maryland} (\bibinfo {year} {1972})\BibitemShut {NoStop}%
\bibitem [{\citenamefont {Misner}(1994)}]{Misner:1994ge}%
  \BibitemOpen
  \bibfield  {author} {\bibinfo {author} {\bibfnamefont {C.~W.}\ \bibnamefont
  {Misner}},\ }\bibfield  {title} {\bibinfo {title} {{The Mixmaster
  cosmological metrics}},\ }in\ \href@noop {} {\emph {\bibinfo {booktitle}
  {{NATO Advanced Research Workshop on Deterministic Chaos in General
  Relativity}}}}\ (\bibinfo {year} {1994})\ \Eprint
  {https://arxiv.org/abs/gr-qc/9405068} {arXiv:gr-qc/9405068} \BibitemShut
  {NoStop}%
\bibitem [{\citenamefont {Damour}\ \emph {et~al.}(2001)\citenamefont {Damour},
  \citenamefont {Henneaux}, \citenamefont {Julia},\ and\ \citenamefont
  {Nicolai}}]{Damour:2001sa}%
  \BibitemOpen
  \bibfield  {author} {\bibinfo {author} {\bibfnamefont {T.}~\bibnamefont
  {Damour}}, \bibinfo {author} {\bibfnamefont {M.}~\bibnamefont {Henneaux}},
  \bibinfo {author} {\bibfnamefont {B.}~\bibnamefont {Julia}},\ and\ \bibinfo
  {author} {\bibfnamefont {H.}~\bibnamefont {Nicolai}},\ }\bibfield  {title}
  {\bibinfo {title} {{Hyperbolic Kac-Moody algebras and chaos in Kaluza-Klein
  models}},\ }\href {https://doi.org/10.1016/S0370-2693(01)00498-1} {\bibfield
  {journal} {\bibinfo  {journal} {Phys. Lett. B}\ }\textbf {\bibinfo {volume}
  {509}},\ \bibinfo {pages} {323} (\bibinfo {year} {2001})},\ \Eprint
  {https://arxiv.org/abs/hep-th/0103094} {arXiv:hep-th/0103094} \BibitemShut
  {NoStop}%
\bibitem [{\citenamefont {Damour}\ \emph {et~al.}(2003)\citenamefont {Damour},
  \citenamefont {Henneaux},\ and\ \citenamefont {Nicolai}}]{Damour:2002et}%
  \BibitemOpen
  \bibfield  {author} {\bibinfo {author} {\bibfnamefont {T.}~\bibnamefont
  {Damour}}, \bibinfo {author} {\bibfnamefont {M.}~\bibnamefont {Henneaux}},\
  and\ \bibinfo {author} {\bibfnamefont {H.}~\bibnamefont {Nicolai}},\
  }\bibfield  {title} {\bibinfo {title} {{Cosmological billiards}},\ }\href
  {https://doi.org/10.1088/0264-9381/20/9/201} {\bibfield  {journal} {\bibinfo
  {journal} {Class. Quant. Grav.}\ }\textbf {\bibinfo {volume} {20}},\ \bibinfo
  {pages} {R145} (\bibinfo {year} {2003})},\ \Eprint
  {https://arxiv.org/abs/hep-th/0212256} {arXiv:hep-th/0212256} \BibitemShut
  {NoStop}%
\bibitem [{\citenamefont {Demaret}\ \emph {et~al.}(1985)\citenamefont
  {Demaret}, \citenamefont {Henneaux},\ and\ \citenamefont
  {Spindel}}]{Demaret:1985jnc}%
  \BibitemOpen
  \bibfield  {author} {\bibinfo {author} {\bibfnamefont {J.}~\bibnamefont
  {Demaret}}, \bibinfo {author} {\bibfnamefont {M.}~\bibnamefont {Henneaux}},\
  and\ \bibinfo {author} {\bibfnamefont {P.}~\bibnamefont {Spindel}},\
  }\bibfield  {title} {\bibinfo {title} {{NONOSCILLATORY BEHAVIOR IN VACUUM
  KALUZA-KLEIN COSMOLOGIES}},\ }\href
  {https://doi.org/10.1016/0370-2693(85)90024-3} {\bibfield  {journal}
  {\bibinfo  {journal} {Phys. Lett. B}\ }\textbf {\bibinfo {volume} {164}},\
  \bibinfo {pages} {27} (\bibinfo {year} {1985})}\BibitemShut {NoStop}%
\bibitem [{\citenamefont {Demaret}\ \emph {et~al.}(1988)\citenamefont
  {Demaret}, \citenamefont {De~Rop},\ and\ \citenamefont
  {Henneaux}}]{Demaret:1988sg}%
  \BibitemOpen
  \bibfield  {author} {\bibinfo {author} {\bibfnamefont {J.}~\bibnamefont
  {Demaret}}, \bibinfo {author} {\bibfnamefont {Y.}~\bibnamefont {De~Rop}},\
  and\ \bibinfo {author} {\bibfnamefont {M.}~\bibnamefont {Henneaux}},\
  }\bibfield  {title} {\bibinfo {title} {{Chaos in Nondiagonal Spatially
  Homogeneous Cosmological Models in Space-time Dimensions $\le$ 10}},\ }\href
  {https://doi.org/10.1016/0370-2693(88)90803-9} {\bibfield  {journal}
  {\bibinfo  {journal} {Phys. Lett. B}\ }\textbf {\bibinfo {volume} {211}},\
  \bibinfo {pages} {37} (\bibinfo {year} {1988})}\BibitemShut {NoStop}%
\bibitem [{\citenamefont {Berger}\ \emph {et~al.}(1998)\citenamefont {Berger},
  \citenamefont {Garfinkle}, \citenamefont {Isenberg}, \citenamefont
  {Moncrief},\ and\ \citenamefont {Weaver}}]{Berger:1998vxa}%
  \BibitemOpen
  \bibfield  {author} {\bibinfo {author} {\bibfnamefont {B.~K.}\ \bibnamefont
  {Berger}}, \bibinfo {author} {\bibfnamefont {D.}~\bibnamefont {Garfinkle}},
  \bibinfo {author} {\bibfnamefont {J.}~\bibnamefont {Isenberg}}, \bibinfo
  {author} {\bibfnamefont {V.}~\bibnamefont {Moncrief}},\ and\ \bibinfo
  {author} {\bibfnamefont {M.}~\bibnamefont {Weaver}},\ }\bibfield  {title}
  {\bibinfo {title} {{The Singularity in generic gravitational collapse is
  space - like, local, and oscillatory}},\ }\href
  {https://doi.org/10.1142/S0217732398001649} {\bibfield  {journal} {\bibinfo
  {journal} {Mod. Phys. Lett. A}\ }\textbf {\bibinfo {volume} {13}},\ \bibinfo
  {pages} {1565} (\bibinfo {year} {1998})},\ \Eprint
  {https://arxiv.org/abs/gr-qc/9805063} {arXiv:gr-qc/9805063} \BibitemShut
  {NoStop}%
\bibitem [{\citenamefont {Garfinkle}(2004)}]{Garfinkle:2003bb}%
  \BibitemOpen
  \bibfield  {author} {\bibinfo {author} {\bibfnamefont {D.}~\bibnamefont
  {Garfinkle}},\ }\bibfield  {title} {\bibinfo {title} {{Numerical simulations
  of generic singuarities}},\ }\href
  {https://doi.org/10.1103/PhysRevLett.93.161101} {\bibfield  {journal}
  {\bibinfo  {journal} {Phys. Rev. Lett.}\ }\textbf {\bibinfo {volume} {93}},\
  \bibinfo {pages} {161101} (\bibinfo {year} {2004})},\ \Eprint
  {https://arxiv.org/abs/gr-qc/0312117} {arXiv:gr-qc/0312117} \BibitemShut
  {NoStop}%
\bibitem [{\citenamefont {Ringstrom}(2001)}]{Ringstrom:2000mk}%
  \BibitemOpen
  \bibfield  {author} {\bibinfo {author} {\bibfnamefont {H.}~\bibnamefont
  {Ringstrom}},\ }\bibfield  {title} {\bibinfo {title} {{The Bianchi IX
  attractor}},\ }\href {https://doi.org/10.1007/PL00001041} {\bibfield
  {journal} {\bibinfo  {journal} {Annales Henri Poincare}\ }\textbf {\bibinfo
  {volume} {2}},\ \bibinfo {pages} {405} (\bibinfo {year} {2001})},\ \Eprint
  {https://arxiv.org/abs/gr-qc/0006035} {arXiv:gr-qc/0006035} \BibitemShut
  {NoStop}%
\bibitem [{\citenamefont {Heinzle}\ \emph {et~al.}(2009)\citenamefont
  {Heinzle}, \citenamefont {Uggla},\ and\ \citenamefont
  {Rohr}}]{Heinzle:2007kv}%
  \BibitemOpen
  \bibfield  {author} {\bibinfo {author} {\bibfnamefont {J.~M.}\ \bibnamefont
  {Heinzle}}, \bibinfo {author} {\bibfnamefont {C.}~\bibnamefont {Uggla}},\
  and\ \bibinfo {author} {\bibfnamefont {N.}~\bibnamefont {Rohr}},\ }\bibfield
  {title} {\bibinfo {title} {{The Cosmological billiard attractor}},\ }\href
  {https://doi.org/10.4310/ATMP.2009.v13.n2.a1} {\bibfield  {journal} {\bibinfo
   {journal} {Adv. Theor. Math. Phys.}\ }\textbf {\bibinfo {volume} {13}},\
  \bibinfo {pages} {293} (\bibinfo {year} {2009})},\ \Eprint
  {https://arxiv.org/abs/gr-qc/0702141} {arXiv:gr-qc/0702141} \BibitemShut
  {NoStop}%
\bibitem [{\citenamefont {Heinzle}\ and\ \citenamefont
  {Uggla}(2009)}]{Heinzle:2009eh}%
  \BibitemOpen
  \bibfield  {author} {\bibinfo {author} {\bibfnamefont {J.~M.}\ \bibnamefont
  {Heinzle}}\ and\ \bibinfo {author} {\bibfnamefont {C.}~\bibnamefont
  {Uggla}},\ }\bibfield  {title} {\bibinfo {title} {{A New proof of the Bianchi
  type IX attractor theorem}},\ }\href
  {https://doi.org/10.1088/0264-9381/26/7/075015} {\bibfield  {journal}
  {\bibinfo  {journal} {Class. Quant. Grav.}\ }\textbf {\bibinfo {volume}
  {26}},\ \bibinfo {pages} {075015} (\bibinfo {year} {2009})},\ \Eprint
  {https://arxiv.org/abs/0901.0806} {arXiv:0901.0806 [gr-qc]} \BibitemShut
  {NoStop}%
\bibitem [{\citenamefont {De~Clerck}\ \emph {et~al.}(2024)\citenamefont
  {De~Clerck}, \citenamefont {Hartnoll},\ and\ \citenamefont
  {Santos}}]{DeClerck:2023fax}%
  \BibitemOpen
  \bibfield  {author} {\bibinfo {author} {\bibfnamefont {M.}~\bibnamefont
  {De~Clerck}}, \bibinfo {author} {\bibfnamefont {S.~A.}\ \bibnamefont
  {Hartnoll}},\ and\ \bibinfo {author} {\bibfnamefont {J.~E.}\ \bibnamefont
  {Santos}},\ }\bibfield  {title} {\bibinfo {title} {{Mixmaster chaos in an AdS
  black hole interior}},\ }\href {https://doi.org/10.1007/JHEP07(2024)202}
  {\bibfield  {journal} {\bibinfo  {journal} {JHEP}\ }\textbf {\bibinfo
  {volume} {07}},\ \bibinfo {pages} {202}},\ \Eprint
  {https://arxiv.org/abs/2312.11622} {arXiv:2312.11622 [hep-th]} \BibitemShut
  {NoStop}%
\bibitem [{\citenamefont {C{\'a}ceres}\ \emph {et~al.}(2026)\citenamefont
  {C{\'a}ceres}, \citenamefont {Murcia}, \citenamefont {Patra},\ and\
  \citenamefont {Pedraza}}]{Caceres:2026mug}%
  \BibitemOpen
  \bibfield  {author} {\bibinfo {author} {\bibfnamefont {E.}~\bibnamefont
  {C{\'a}ceres}}, \bibinfo {author} {\bibfnamefont {{\'A}.~J.}\ \bibnamefont
  {Murcia}}, \bibinfo {author} {\bibfnamefont {A.~K.}\ \bibnamefont {Patra}},\
  and\ \bibinfo {author} {\bibfnamefont {J.~F.}\ \bibnamefont {Pedraza}},\
  }\bibfield  {title} {\bibinfo {title} {{Higher-dimensional BKL dynamics in
  AdS black holes}},\ }\href {https://doi.org/10.1007/JHEP08(2026)164}
  {\bibfield  {journal} {\bibinfo  {journal} {JHEP}\ }\textbf {\bibinfo
  {volume} {08}},\ \bibinfo {pages} {164}},\ \Eprint
  {https://arxiv.org/abs/2603.06560} {arXiv:2603.06560 [hep-th]} \BibitemShut
  {NoStop}%
\bibitem [{Note1()}]{Note1}%
  \BibitemOpen
  \bibinfo {note} {These examples use negative squared mass for the vector to
  ensure absence of inner horizons. The models satisfy the AdS BF bound but
  violate the CFT unitarity bound~\cite {DeClerck:2023fax,Caceres:2026mug}. It
  is expected that extensions exist that avoid this problem~\cite
  {DeClerck:2023fax}, but these have not yet been constructed. A further
  qualification concerns longitudinal modes, which the ansatz suppresses. For
  positive mass squared, exciting these modes can open the billiard and
  terminate the oscillations~\cite {Henneaux:2022ijt}.}\BibitemShut {Stop}%
\bibitem [{\citenamefont {Misner}(1969)}]{Misner:1969hg}%
  \BibitemOpen
  \bibfield  {author} {\bibinfo {author} {\bibfnamefont {C.~W.}\ \bibnamefont
  {Misner}},\ }\bibfield  {title} {\bibinfo {title} {{Mixmaster universe}},\
  }\href {https://doi.org/10.1103/PhysRevLett.22.1071} {\bibfield  {journal}
  {\bibinfo  {journal} {Phys. Rev. Lett.}\ }\textbf {\bibinfo {volume} {22}},\
  \bibinfo {pages} {1071} (\bibinfo {year} {1969})}\BibitemShut {NoStop}%
\bibitem [{\citenamefont {Siklos}(1978)}]{Siklos:1978}%
  \BibitemOpen
  \bibfield  {author} {\bibinfo {author} {\bibfnamefont {S.~T.~C.}\
  \bibnamefont {Siklos}},\ }\bibfield  {title} {\bibinfo {title} {{Occurrence
  of whimper singularities}},\ }\href {https://doi.org/10.1007/BF01614223}
  {\bibfield  {journal} {\bibinfo  {journal} {Commun. Math. Phys.}\ }\textbf
  {\bibinfo {volume} {58}},\ \bibinfo {pages} {255} (\bibinfo {year}
  {1978})}\BibitemShut {NoStop}%
\bibitem [{\citenamefont {Chrusciel}\ and\ \citenamefont
  {Rendall}(1995)}]{Chrusciel:1994be}%
  \BibitemOpen
  \bibfield  {author} {\bibinfo {author} {\bibfnamefont {P.~T.}\ \bibnamefont
  {Chrusciel}}\ and\ \bibinfo {author} {\bibfnamefont {A.~D.}\ \bibnamefont
  {Rendall}},\ }\bibfield  {title} {\bibinfo {title} {{Strong cosmic censorship
  in vacuum space-times with compact, locally homogeneous Cauchy surfaces}},\
  }\href {https://doi.org/10.1006/aphy.1995.1084} {\bibfield  {journal}
  {\bibinfo  {journal} {Annals Phys.}\ }\textbf {\bibinfo {volume} {242}},\
  \bibinfo {pages} {349} (\bibinfo {year} {1995})},\ \Eprint
  {https://arxiv.org/abs/gr-qc/9410040} {arXiv:gr-qc/9410040} \BibitemShut
  {NoStop}%
\bibitem [{\citenamefont {Ringstrom}(2000)}]{Ringstrom:1999hy}%
  \BibitemOpen
  \bibfield  {author} {\bibinfo {author} {\bibfnamefont {H.}~\bibnamefont
  {Ringstrom}},\ }\bibfield  {title} {\bibinfo {title} {{Curvature blowup in
  Bianchi VIII and IX vacuum space-times}},\ }\href
  {https://doi.org/10.1088/0264-9381/17/4/301} {\bibfield  {journal} {\bibinfo
  {journal} {Class. Quant. Grav.}\ }\textbf {\bibinfo {volume} {17}},\ \bibinfo
  {pages} {713} (\bibinfo {year} {2000})},\ \Eprint
  {https://arxiv.org/abs/gr-qc/9911115} {arXiv:gr-qc/9911115} \BibitemShut
  {NoStop}%
\bibitem [{\citenamefont {Hennigar}\ and\ \citenamefont
  {Patra}(2026)}]{ForthcomingBKL}%
  \BibitemOpen
  \bibfield  {author} {\bibinfo {author} {\bibfnamefont {R.~A.}\ \bibnamefont
  {Hennigar}}\ and\ \bibinfo {author} {\bibfnamefont {A.~K.}\ \bibnamefont
  {Patra}},\ }\bibfield  {title} {\bibinfo {title} {{AdS black holes with BKL
  interiors}}} (\bibinfo {year} {2026}),\ \bibinfo {note}
  {forthcoming}\BibitemShut {NoStop}%
\bibitem [{\citenamefont {Murata}\ \emph {et~al.}(2009)\citenamefont {Murata},
  \citenamefont {Nishioka},\ and\ \citenamefont {Tanahashi}}]{Murata:2009jt}%
  \BibitemOpen
  \bibfield  {author} {\bibinfo {author} {\bibfnamefont {K.}~\bibnamefont
  {Murata}}, \bibinfo {author} {\bibfnamefont {T.}~\bibnamefont {Nishioka}},\
  and\ \bibinfo {author} {\bibfnamefont {N.}~\bibnamefont {Tanahashi}},\
  }\bibfield  {title} {\bibinfo {title} {{Warped AdS(5) Black Holes and Dual
  CFTs}},\ }\href {https://doi.org/10.1143/PTP.121.941} {\bibfield  {journal}
  {\bibinfo  {journal} {Prog. Theor. Phys.}\ }\textbf {\bibinfo {volume}
  {121}},\ \bibinfo {pages} {941} (\bibinfo {year} {2009})},\ \Eprint
  {https://arxiv.org/abs/0901.2574} {arXiv:0901.2574 [hep-th]} \BibitemShut
  {NoStop}%
\bibitem [{\citenamefont {Brihaye}\ \emph {et~al.}(2009)\citenamefont
  {Brihaye}, \citenamefont {Kunz},\ and\ \citenamefont
  {Radu}}]{Brihaye:2009dm}%
  \BibitemOpen
  \bibfield  {author} {\bibinfo {author} {\bibfnamefont {Y.}~\bibnamefont
  {Brihaye}}, \bibinfo {author} {\bibfnamefont {J.}~\bibnamefont {Kunz}},\ and\
  \bibinfo {author} {\bibfnamefont {E.}~\bibnamefont {Radu}},\ }\bibfield
  {title} {\bibinfo {title} {{From black strings to black holes: Nuttier and
  squashed AdS(5) solutions}},\ }\href
  {https://doi.org/10.1088/1126-6708/2009/08/025} {\bibfield  {journal}
  {\bibinfo  {journal} {JHEP}\ }\textbf {\bibinfo {volume} {08}},\ \bibinfo
  {pages} {025}},\ \Eprint {https://arxiv.org/abs/0904.1566} {arXiv:0904.1566
  [gr-qc]} \BibitemShut {NoStop}%
\bibitem [{\citenamefont {Hawking}\ \emph {et~al.}(1999)\citenamefont
  {Hawking}, \citenamefont {Hunter},\ and\ \citenamefont
  {Taylor}}]{Hawking:1998kw}%
  \BibitemOpen
  \bibfield  {author} {\bibinfo {author} {\bibfnamefont {S.~W.}\ \bibnamefont
  {Hawking}}, \bibinfo {author} {\bibfnamefont {C.~J.}\ \bibnamefont
  {Hunter}},\ and\ \bibinfo {author} {\bibfnamefont {M.}~\bibnamefont
  {Taylor}},\ }\bibfield  {title} {\bibinfo {title} {{Rotation and the AdS /
  CFT correspondence}},\ }\href {https://doi.org/10.1103/PhysRevD.59.064005}
  {\bibfield  {journal} {\bibinfo  {journal} {Phys. Rev. D}\ }\textbf {\bibinfo
  {volume} {59}},\ \bibinfo {pages} {064005} (\bibinfo {year} {1999})},\
  \Eprint {https://arxiv.org/abs/hep-th/9811056} {arXiv:hep-th/9811056}
  \BibitemShut {NoStop}%
\bibitem [{\citenamefont {Ishii}\ and\ \citenamefont
  {Murata}(2019)}]{Ishii:2018oms}%
  \BibitemOpen
  \bibfield  {author} {\bibinfo {author} {\bibfnamefont {T.}~\bibnamefont
  {Ishii}}\ and\ \bibinfo {author} {\bibfnamefont {K.}~\bibnamefont {Murata}},\
  }\bibfield  {title} {\bibinfo {title} {{Black resonators and geons in
  AdS5}},\ }\href {https://doi.org/10.1088/1361-6382/ab1d76} {\bibfield
  {journal} {\bibinfo  {journal} {Class. Quant. Grav.}\ }\textbf {\bibinfo
  {volume} {36}},\ \bibinfo {pages} {125011} (\bibinfo {year} {2019})},\
  \Eprint {https://arxiv.org/abs/1810.11089} {arXiv:1810.11089 [hep-th]}
  \BibitemShut {NoStop}%
\bibitem [{Note2()}]{Note2}%
  \BibitemOpen
  \bibinfo {note} {Note that diagonalizing $H_{ab}$ at the horizon does not
  mean $h_{ab}$ will be diagonal off the horizon.}\BibitemShut {Stop}%
\bibitem [{\citenamefont {de~Buyl}\ \emph {et~al.}(2003)\citenamefont
  {de~Buyl}, \citenamefont {Pinardi},\ and\ \citenamefont
  {Schomblond}}]{deBuyl:2003ub}%
  \BibitemOpen
  \bibfield  {author} {\bibinfo {author} {\bibfnamefont {S.}~\bibnamefont
  {de~Buyl}}, \bibinfo {author} {\bibfnamefont {G.}~\bibnamefont {Pinardi}},\
  and\ \bibinfo {author} {\bibfnamefont {C.}~\bibnamefont {Schomblond}},\
  }\bibfield  {title} {\bibinfo {title} {{Einstein billiards and spatially
  homogeneous cosmological models}},\ }\href
  {https://doi.org/10.1088/0264-9381/20/23/012} {\bibfield  {journal} {\bibinfo
   {journal} {Class. Quant. Grav.}\ }\textbf {\bibinfo {volume} {20}},\
  \bibinfo {pages} {5141} (\bibinfo {year} {2003})},\ \Eprint
  {https://arxiv.org/abs/hep-th/0306280} {arXiv:hep-th/0306280} \BibitemShut
  {NoStop}%
\bibitem [{\citenamefont {Hartnoll}\ and\ \citenamefont
  {Yang}(2025)}]{Hartnoll:2025hly}%
  \BibitemOpen
  \bibfield  {author} {\bibinfo {author} {\bibfnamefont {S.~A.}\ \bibnamefont
  {Hartnoll}}\ and\ \bibinfo {author} {\bibfnamefont {M.}~\bibnamefont
  {Yang}},\ }\bibfield  {title} {\bibinfo {title} {{The conformal primon gas at
  the end of time}},\ }\href {https://doi.org/10.1007/JHEP07(2025)281}
  {\bibfield  {journal} {\bibinfo  {journal} {JHEP}\ }\textbf {\bibinfo
  {volume} {07}},\ \bibinfo {pages} {281}},\ \Eprint
  {https://arxiv.org/abs/2502.02661} {arXiv:2502.02661 [hep-th]} \BibitemShut
  {NoStop}%
\bibitem [{\citenamefont {De~Clerck}\ \emph {et~al.}(2025)\citenamefont
  {De~Clerck}, \citenamefont {Hartnoll},\ and\ \citenamefont
  {Yang}}]{DeClerck:2025mem}%
  \BibitemOpen
  \bibfield  {author} {\bibinfo {author} {\bibfnamefont {M.}~\bibnamefont
  {De~Clerck}}, \bibinfo {author} {\bibfnamefont {S.~A.}\ \bibnamefont
  {Hartnoll}},\ and\ \bibinfo {author} {\bibfnamefont {M.}~\bibnamefont
  {Yang}},\ }\bibfield  {title} {\bibinfo {title} {{Wheeler-DeWitt
  wavefunctions for 5d BKL dynamics, automorphic L-functions and complex primon
  gases}},\ }\href {https://doi.org/10.1007/JHEP11(2025)160} {\bibfield
  {journal} {\bibinfo  {journal} {JHEP}\ }\textbf {\bibinfo {volume} {11}},\
  \bibinfo {pages} {160}},\ \Eprint {https://arxiv.org/abs/2507.08788}
  {arXiv:2507.08788 [hep-th]} \BibitemShut {NoStop}%
\bibitem [{\citenamefont {Bueno}\ \emph
  {et~al.}(2024{\natexlab{a}})\citenamefont {Bueno}, \citenamefont {Cano},\
  and\ \citenamefont {Hennigar}}]{Bueno:2024fzg}%
  \BibitemOpen
  \bibfield  {author} {\bibinfo {author} {\bibfnamefont {P.}~\bibnamefont
  {Bueno}}, \bibinfo {author} {\bibfnamefont {P.~A.}\ \bibnamefont {Cano}},\
  and\ \bibinfo {author} {\bibfnamefont {R.~A.}\ \bibnamefont {Hennigar}},\
  }\bibfield  {title} {\bibinfo {title} {{Kasner epochs, eras and eons}},\
  }\href {https://doi.org/10.1103/PhysRevD.110.L041503} {\bibfield  {journal}
  {\bibinfo  {journal} {Phys. Rev. D}\ }\textbf {\bibinfo {volume} {110}},\
  \bibinfo {pages} {L041503} (\bibinfo {year} {2024}{\natexlab{a}})},\ \Eprint
  {https://arxiv.org/abs/2402.14912} {arXiv:2402.14912 [gr-qc]} \BibitemShut
  {NoStop}%
\bibitem [{\citenamefont {C{\'a}ceres}\ \emph {et~al.}(2024)\citenamefont
  {C{\'a}ceres}, \citenamefont {Murcia}, \citenamefont {Patra},\ and\
  \citenamefont {Pedraza}}]{Caceres:2024edr}%
  \BibitemOpen
  \bibfield  {author} {\bibinfo {author} {\bibfnamefont {E.}~\bibnamefont
  {C{\'a}ceres}}, \bibinfo {author} {\bibfnamefont {{\'A}.~J.}\ \bibnamefont
  {Murcia}}, \bibinfo {author} {\bibfnamefont {A.~K.}\ \bibnamefont {Patra}},\
  and\ \bibinfo {author} {\bibfnamefont {J.~F.}\ \bibnamefont {Pedraza}},\
  }\bibfield  {title} {\bibinfo {title} {{Kasner eons with matter: holographic
  excursions to the black hole singularity}},\ }\href
  {https://doi.org/10.1007/JHEP12(2024)077} {\bibfield  {journal} {\bibinfo
  {journal} {JHEP}\ }\textbf {\bibinfo {volume} {12}},\ \bibinfo {pages}
  {077}},\ \Eprint {https://arxiv.org/abs/2408.14535} {arXiv:2408.14535
  [hep-th]} \BibitemShut {NoStop}%
\bibitem [{\citenamefont {Bueno}\ \emph
  {et~al.}(2024{\natexlab{b}})\citenamefont {Bueno}, \citenamefont {Cano},
  \citenamefont {Hennigar},\ and\ \citenamefont {Li}}]{Bueno:2024qhh}%
  \BibitemOpen
  \bibfield  {author} {\bibinfo {author} {\bibfnamefont {P.}~\bibnamefont
  {Bueno}}, \bibinfo {author} {\bibfnamefont {P.~A.}\ \bibnamefont {Cano}},
  \bibinfo {author} {\bibfnamefont {R.~A.}\ \bibnamefont {Hennigar}},\ and\
  \bibinfo {author} {\bibfnamefont {M.-D.}\ \bibnamefont {Li}},\ }\bibfield
  {title} {\bibinfo {title} {{Kasner eons in Lovelock black holes}},\ }\href
  {https://doi.org/10.1103/PhysRevD.110.124015} {\bibfield  {journal} {\bibinfo
   {journal} {Phys. Rev. D}\ }\textbf {\bibinfo {volume} {110}},\ \bibinfo
  {pages} {124015} (\bibinfo {year} {2024}{\natexlab{b}})},\ \Eprint
  {https://arxiv.org/abs/2409.00648} {arXiv:2409.00648 [hep-th]} \BibitemShut
  {NoStop}%
\bibitem [{\citenamefont {Fidkowski}\ \emph {et~al.}(2004)\citenamefont
  {Fidkowski}, \citenamefont {Hubeny}, \citenamefont {Kleban},\ and\
  \citenamefont {Shenker}}]{Fidkowski:2003nf}%
  \BibitemOpen
  \bibfield  {author} {\bibinfo {author} {\bibfnamefont {L.}~\bibnamefont
  {Fidkowski}}, \bibinfo {author} {\bibfnamefont {V.}~\bibnamefont {Hubeny}},
  \bibinfo {author} {\bibfnamefont {M.}~\bibnamefont {Kleban}},\ and\ \bibinfo
  {author} {\bibfnamefont {S.}~\bibnamefont {Shenker}},\ }\bibfield  {title}
  {\bibinfo {title} {{The Black hole singularity in AdS / CFT}},\ }\href
  {https://doi.org/10.1088/1126-6708/2004/02/014} {\bibfield  {journal}
  {\bibinfo  {journal} {JHEP}\ }\textbf {\bibinfo {volume} {02}},\ \bibinfo
  {pages} {014}},\ \Eprint {https://arxiv.org/abs/hep-th/0306170}
  {arXiv:hep-th/0306170} \BibitemShut {NoStop}%
\bibitem [{\citenamefont {Festuccia}\ and\ \citenamefont
  {Liu}(2006)}]{Festuccia:2005pi}%
  \BibitemOpen
  \bibfield  {author} {\bibinfo {author} {\bibfnamefont {G.}~\bibnamefont
  {Festuccia}}\ and\ \bibinfo {author} {\bibfnamefont {H.}~\bibnamefont
  {Liu}},\ }\bibfield  {title} {\bibinfo {title} {{Excursions beyond the
  horizon: Black hole singularities in Yang-Mills theories. I.}},\ }\href
  {https://doi.org/10.1088/1126-6708/2006/04/044} {\bibfield  {journal}
  {\bibinfo  {journal} {JHEP}\ }\textbf {\bibinfo {volume} {04}},\ \bibinfo
  {pages} {044}},\ \Eprint {https://arxiv.org/abs/hep-th/0506202}
  {arXiv:hep-th/0506202} \BibitemShut {NoStop}%
\bibitem [{\citenamefont {Frenkel}\ \emph {et~al.}(2020)\citenamefont
  {Frenkel}, \citenamefont {Hartnoll}, \citenamefont {Kruthoff},\ and\
  \citenamefont {Shi}}]{Frenkel:2020ysx}%
  \BibitemOpen
  \bibfield  {author} {\bibinfo {author} {\bibfnamefont {A.}~\bibnamefont
  {Frenkel}}, \bibinfo {author} {\bibfnamefont {S.~A.}\ \bibnamefont
  {Hartnoll}}, \bibinfo {author} {\bibfnamefont {J.}~\bibnamefont {Kruthoff}},\
  and\ \bibinfo {author} {\bibfnamefont {Z.~D.}\ \bibnamefont {Shi}},\
  }\bibfield  {title} {\bibinfo {title} {{Holographic flows from CFT to the
  Kasner universe}},\ }\href {https://doi.org/10.1007/JHEP08(2020)003}
  {\bibfield  {journal} {\bibinfo  {journal} {JHEP}\ }\textbf {\bibinfo
  {volume} {08}},\ \bibinfo {pages} {003}},\ \Eprint
  {https://arxiv.org/abs/2004.01192} {arXiv:2004.01192 [hep-th]} \BibitemShut
  {NoStop}%
\bibitem [{\citenamefont {Hartnoll}\ and\ \citenamefont
  {Zhiboedov}(2026)}]{Hartnoll:2026vhu}%
  \BibitemOpen
  \bibfield  {author} {\bibinfo {author} {\bibfnamefont {S.~A.}\ \bibnamefont
  {Hartnoll}}\ and\ \bibinfo {author} {\bibfnamefont {A.}~\bibnamefont
  {Zhiboedov}},\ }\bibfield  {title} {\bibinfo {title} {{Can one hear the shape
  of a black hole singularity?}},\ }\href@noop {} {\  (\bibinfo {year}
  {2026})},\ \Eprint {https://arxiv.org/abs/2609.07514} {arXiv:2609.07514
  [hep-th]} \BibitemShut {NoStop}%
\bibitem [{\citenamefont {Xiao}\ \emph {et~al.}(2026)\citenamefont {Xiao},
  \citenamefont {Nian},\ and\ \citenamefont {Li}}]{Xiao:2026pir}%
  \BibitemOpen
  \bibfield  {author} {\bibinfo {author} {\bibfnamefont {Z.-Q.}\ \bibnamefont
  {Xiao}}, \bibinfo {author} {\bibfnamefont {J.}~\bibnamefont {Nian}},\ and\
  \bibinfo {author} {\bibfnamefont {L.}~\bibnamefont {Li}},\ }\bibfield
  {title} {\bibinfo {title} {{Channel-Resolved High-Overtone Quasinormal Modes
  Decode Kasner Scaling inside Black Holes}},\ }\href@noop {} {\  (\bibinfo
  {year} {2026})},\ \Eprint {https://arxiv.org/abs/2609.17426}
  {arXiv:2609.17426 [hep-th]} \BibitemShut {NoStop}%
\bibitem [{\citenamefont {Coviello}\ \emph
  {et~al.}(2026{\natexlab{a}})\citenamefont {Coviello}, \citenamefont {Gupta},
  \citenamefont {Hennigar}, \citenamefont {Shi},\ and\ \citenamefont
  {Svesko}}]{Coviello:2026tol}%
  \BibitemOpen
  \bibfield  {author} {\bibinfo {author} {\bibfnamefont {C.}~\bibnamefont
  {Coviello}}, \bibinfo {author} {\bibfnamefont {A.}~\bibnamefont {Gupta}},
  \bibinfo {author} {\bibfnamefont {R.~A.}\ \bibnamefont {Hennigar}}, \bibinfo
  {author} {\bibfnamefont {K.}~\bibnamefont {Shi}},\ and\ \bibinfo {author}
  {\bibfnamefont {A.}~\bibnamefont {Svesko}},\ }\bibfield  {title} {\bibinfo
  {title} {{Looking inside a quantum black hole}},\ }\href@noop {} {\
  (\bibinfo {year} {2026}{\natexlab{a}})},\ \Eprint
  {https://arxiv.org/abs/2609.20788} {arXiv:2609.20788 [hep-th]} \BibitemShut
  {NoStop}%
\bibitem [{\citenamefont {Coviello}\ \emph
  {et~al.}(2026{\natexlab{b}})\citenamefont {Coviello}, \citenamefont {Gupta},
  \citenamefont {Hennigar}, \citenamefont {Shi},\ and\ \citenamefont
  {Svesko}}]{Coviello:2026hdc}%
  \BibitemOpen
  \bibfield  {author} {\bibinfo {author} {\bibfnamefont {C.}~\bibnamefont
  {Coviello}}, \bibinfo {author} {\bibfnamefont {A.}~\bibnamefont {Gupta}},
  \bibinfo {author} {\bibfnamefont {R.~A.}\ \bibnamefont {Hennigar}}, \bibinfo
  {author} {\bibfnamefont {K.}~\bibnamefont {Shi}},\ and\ \bibinfo {author}
  {\bibfnamefont {A.}~\bibnamefont {Svesko}},\ }\bibfield  {title} {\bibinfo
  {title} {{Quasinormal modes and quantum black hole interiors}},\ }\href@noop
  {} {\  (\bibinfo {year} {2026}{\natexlab{b}})},\ \Eprint
  {https://arxiv.org/abs/2609.26724} {arXiv:2609.26724 [hep-th]} \BibitemShut
  {NoStop}%
\bibitem [{\citenamefont {{\v{C}}eplak}\ \emph {et~al.}(2024)\citenamefont
  {{\v{C}}eplak}, \citenamefont {Liu}, \citenamefont {Parnachev},\ and\
  \citenamefont {Valach}}]{Ceplak:2024bja}%
  \BibitemOpen
  \bibfield  {author} {\bibinfo {author} {\bibfnamefont {N.}~\bibnamefont
  {{\v{C}}eplak}}, \bibinfo {author} {\bibfnamefont {H.}~\bibnamefont {Liu}},
  \bibinfo {author} {\bibfnamefont {A.}~\bibnamefont {Parnachev}},\ and\
  \bibinfo {author} {\bibfnamefont {S.}~\bibnamefont {Valach}},\ }\bibfield
  {title} {\bibinfo {title} {{Black hole singularity from OPE}},\ }\href
  {https://doi.org/10.1007/JHEP10(2024)105} {\bibfield  {journal} {\bibinfo
  {journal} {JHEP}\ }\textbf {\bibinfo {volume} {10}},\ \bibinfo {pages}
  {105}},\ \Eprint {https://arxiv.org/abs/2404.17286} {arXiv:2404.17286
  [hep-th]} \BibitemShut {NoStop}%
\bibitem [{\citenamefont {{\v{C}}eplak}\ and\ \citenamefont
  {Valach}(2026)}]{Ceplak:2026ard}%
  \BibitemOpen
  \bibfield  {author} {\bibinfo {author} {\bibfnamefont {N.}~\bibnamefont
  {{\v{C}}eplak}}\ and\ \bibinfo {author} {\bibfnamefont {S.}~\bibnamefont
  {Valach}},\ }\bibfield  {title} {\bibinfo {title} {{Where's Kasner? OPE
  Signature of Kasner Exponents}},\ }\href@noop {} {\  (\bibinfo {year}
  {2026})},\ \Eprint {https://arxiv.org/abs/2609.23776} {arXiv:2609.23776
  [hep-th]} \BibitemShut {NoStop}%
\bibitem [{\citenamefont {Hartnoll}\ \emph {et~al.}(2021)\citenamefont
  {Hartnoll}, \citenamefont {Horowitz}, \citenamefont {Kruthoff},\ and\
  \citenamefont {Santos}}]{Hartnoll:2020fhc}%
  \BibitemOpen
  \bibfield  {author} {\bibinfo {author} {\bibfnamefont {S.~A.}\ \bibnamefont
  {Hartnoll}}, \bibinfo {author} {\bibfnamefont {G.~T.}\ \bibnamefont
  {Horowitz}}, \bibinfo {author} {\bibfnamefont {J.}~\bibnamefont {Kruthoff}},\
  and\ \bibinfo {author} {\bibfnamefont {J.~E.}\ \bibnamefont {Santos}},\
  }\bibfield  {title} {\bibinfo {title} {{Diving into a holographic
  superconductor}},\ }\href {https://doi.org/10.21468/SciPostPhys.10.1.009}
  {\bibfield  {journal} {\bibinfo  {journal} {SciPost Phys.}\ }\textbf
  {\bibinfo {volume} {10}},\ \bibinfo {pages} {009} (\bibinfo {year} {2021})},\
  \Eprint {https://arxiv.org/abs/2008.12786} {arXiv:2008.12786 [hep-th]}
  \BibitemShut {NoStop}%
\bibitem [{\citenamefont {Cai}\ \emph {et~al.}(2021)\citenamefont {Cai},
  \citenamefont {Li},\ and\ \citenamefont {Yang}}]{Cai:2020wrp}%
  \BibitemOpen
  \bibfield  {author} {\bibinfo {author} {\bibfnamefont {R.-G.}\ \bibnamefont
  {Cai}}, \bibinfo {author} {\bibfnamefont {L.}~\bibnamefont {Li}},\ and\
  \bibinfo {author} {\bibfnamefont {R.-Q.}\ \bibnamefont {Yang}},\ }\bibfield
  {title} {\bibinfo {title} {{No Inner-Horizon Theorem for Black Holes with
  Charged Scalar Hairs}},\ }\href {https://doi.org/10.1007/JHEP03(2021)263}
  {\bibfield  {journal} {\bibinfo  {journal} {JHEP}\ }\textbf {\bibinfo
  {volume} {03}},\ \bibinfo {pages} {263}},\ \Eprint
  {https://arxiv.org/abs/2009.05520} {arXiv:2009.05520 [gr-qc]} \BibitemShut
  {NoStop}%
\bibitem [{\citenamefont {An}\ \emph {et~al.}(2021)\citenamefont {An},
  \citenamefont {Li},\ and\ \citenamefont {Yang}}]{An:2021plu}%
  \BibitemOpen
  \bibfield  {author} {\bibinfo {author} {\bibfnamefont {Y.-S.}\ \bibnamefont
  {An}}, \bibinfo {author} {\bibfnamefont {L.}~\bibnamefont {Li}},\ and\
  \bibinfo {author} {\bibfnamefont {F.-G.}\ \bibnamefont {Yang}},\ }\bibfield
  {title} {\bibinfo {title} {{No Cauchy horizon theorem for nonlinear
  electrodynamics black holes with charged scalar hairs}},\ }\href
  {https://doi.org/10.1103/PhysRevD.104.024040} {\bibfield  {journal} {\bibinfo
   {journal} {Phys. Rev. D}\ }\textbf {\bibinfo {volume} {104}},\ \bibinfo
  {pages} {024040} (\bibinfo {year} {2021})},\ \Eprint
  {https://arxiv.org/abs/2106.01069} {arXiv:2106.01069 [gr-qc]} \BibitemShut
  {NoStop}%
\bibitem [{\citenamefont {Dias}\ \emph {et~al.}(2021)\citenamefont {Dias},
  \citenamefont {Horowitz},\ and\ \citenamefont {Santos}}]{Dias:2021afz}%
  \BibitemOpen
  \bibfield  {author} {\bibinfo {author} {\bibfnamefont {O.~J.~C.}\
  \bibnamefont {Dias}}, \bibinfo {author} {\bibfnamefont {G.~T.}\ \bibnamefont
  {Horowitz}},\ and\ \bibinfo {author} {\bibfnamefont {J.~E.}\ \bibnamefont
  {Santos}},\ }\bibfield  {title} {\bibinfo {title} {{Inside an asymptotically
  flat hairy black hole}},\ }\href {https://doi.org/10.1007/JHEP12(2021)179}
  {\bibfield  {journal} {\bibinfo  {journal} {JHEP}\ }\textbf {\bibinfo
  {volume} {12}},\ \bibinfo {pages} {179}},\ \Eprint
  {https://arxiv.org/abs/2110.06225} {arXiv:2110.06225 [hep-th]} \BibitemShut
  {NoStop}%
\bibitem [{\citenamefont {Belinski}\ and\ \citenamefont
  {Henneaux}(2017)}]{Belinski:2017fas}%
  \BibitemOpen
  \bibfield  {author} {\bibinfo {author} {\bibfnamefont {V.}~\bibnamefont
  {Belinski}}\ and\ \bibinfo {author} {\bibfnamefont {M.}~\bibnamefont
  {Henneaux}},\ }\href {https://doi.org/10.1017/9781107239333} {\emph {\bibinfo
  {title} {{The Cosmological Singularity}}}},\ Cambridge Monogr.Math.Phys.\
  (\bibinfo  {publisher} {Cambridge Univ. Pr.},\ \bibinfo {address}
  {Cambridge},\ \bibinfo {year} {2017})\BibitemShut {NoStop}%
\bibitem [{\citenamefont {Henneaux}(2022)}]{Henneaux:2022ijt}%
  \BibitemOpen
  \bibfield  {author} {\bibinfo {author} {\bibfnamefont {M.}~\bibnamefont
  {Henneaux}},\ }\bibfield  {title} {\bibinfo {title} {{The final Kasner regime
  inside black holes with scalar or vector hair}},\ }\href
  {https://doi.org/10.1007/JHEP03(2022)062} {\bibfield  {journal} {\bibinfo
  {journal} {JHEP}\ }\textbf {\bibinfo {volume} {03}},\ \bibinfo {pages}
  {062}},\ \Eprint {https://arxiv.org/abs/2202.04155} {arXiv:2202.04155
  [hep-th]} \BibitemShut {NoStop}%
\end{thebibliography}%

\end{document}